\documentclass[12pt]{article}
\pdfoutput=1
\usepackage{jheppub}
\usepackage[utf8]{inputenc}
\usepackage{verbatim}
\usepackage{amsmath}
\usepackage{amssymb}
\usepackage{amsthm}
\usepackage{amsfonts}
\usepackage{hyperref}
\usepackage{physics}
\usepackage{float}
\usepackage{xcolor}
\newcommand{\be}{\begin{equation}}
\newcommand{\ee}{\end{equation}}
\newcommand{\ran}{\rangle}
\newcommand{\lan}{\langle}
\newcommand{\mO}{\mathcal{O}}
\newcommand{\ol}{\overline}
\newcommand{\wt}{\widetilde}
\newcommand{\bi}{\begin{itemize}}
\newcommand{\ei}{\end{itemize}}
\newcommand{\bfig}{\begin{figure}\begin{center}}
\newcommand{\efig}{\end{center}\end{figure}}

\newcommand{\CRT}{\mathcal{CRT}}

\usepackage{graphicx}

\DeclareMathOperator{\arcsinh}{arcsinh}

\begin{document}
\title{Quantum mechanics and observers for gravity in a closed universe}
\author[a]{Daniel Harlow}
\author[b]{Mykhaylo Usatyuk}
\author[a,b]{Ying Zhao}
\affiliation[a]{Center for Theoretical Physics\\ Massachusetts Institute of Technology, Cambridge, MA 02139, USA}
\affiliation[b]{Kavli Institute for Theoretical Physics, Santa Barbara, CA 93106, USA}
\emailAdd{harlow@mit.edu,\ musatyuk@kitp.ucsb.edu,\   zhaoying@mit.edu}
\abstract{Recent arguments based on the quantum extremal surface formula or the gravitational path integral have given fairly compelling evidence that the Hilbert space of quantum gravity in a closed universe is one-dimensional and real.  How can this be consistent with the complexity of our own experiences?  In this paper we propose that the experiences of any observer $Ob$ in a closed universe can be approximately described by a quantum mechanical theory with a Hilbert space whose dimension is roughly $e^{S_{Ob}}$, where $S_{Ob}$ is the number of degrees of freedom of $Ob$.  Moreover we argue that the errors in this description are exponentially small in $S_{Ob}$. We give evidence for this proposal using the gravitational path integral and the coding interpretation of holography, and we explain how similar effects arise in black hole physics in appropriate circumstances.}

\maketitle

\section{Introduction}
In recent years there has been remarkable progress on the black hole information problem.  Starting with the initial breakthrough \cite{Penington:2019npb,Almheiri:2019psf}, it was understood that semiclassical methods such as the quantum extremal surface formula and the gravitational path integral can give compelling evidence for the unitarity of black hole evaporation \cite{Almheiri:2019hni,Almheiri:2019yqk,Penington:2019kki,Almheiri:2019qdq,Almheiri:2020cfm}.  These calculations were then given a microscopic interpretation in the language of ``non-isometric codes'', where the Hilbert space of gravitational effective field theory is mapped into some fundamental Hilbert space by a linear but not necessarily isometric encoding map \cite{Akers:2021fut,Akers:2022qdl,Kim:2022pfp,Kar:2022qkf,Chandra:2023dgq}.

This progress however has not come without cost: for sufficiently old black holes the fundamental description of interior observables is likely nonlinear \cite{Marolf:2012xe,Almheiri:2013hfa,Papadodimas:2015jra,Kourkoulou:2017zaj,Almheiri:2018xdw,Akers:2022qdl}.  This nonlinearity calls into question the general validity of quantum mechanics: if the fundamental rule for computing expectation values of observables is something other than ``take the expectation value of a linear self-adjoint operator'', then what exactly are we doing?  In \cite{Akers:2022qdl} it was argued that linear self-adjoint operators in the fundamental description only define observables for observers who are external to the system. Inside the system, and in particular behind a black hole horizon, there is some inherent ambiguity in the notions of ``observer'' and ``observable'', and the fundamental description of an observable can and in general must be something more complicated than a linear self-adjoint operator.  Moreover rules were given for computing interior observables in the fundamental description that agree with the semiclassical description up to ambiguities which are exponentially small in the entropy of the black hole at the time the observer jumps in, which is perhaps the most we should ask for.

It is natural to try to apply this recent progress on black holes to the problem of quantum cosmology.  Unfortunately doing so in the most straightforward way leads to a rather shocking conclusion: the Hilbert space of quantum gravity in a closed universe is one-dimensional \cite{Almheiri:2019hni,Penington:2019kki,Marolf:2020xie,McNamara:2020uza,Usatyuk:2024mzs}. How can it make sense for the Hilbert space of the entire universe, with all its rich complexity, to be one dimensional?  In particular if we are observers living in a closed universe, what calculation are we supposed to do involving this one-dimensional Hilbert space that tells us the likely outcome of an experiment of our choosing?  The main result of this paper is a proposal for how to do this. The essential point is that we must include the observer who is doing the measurement in a systematic way in order to get reasonable answers.  Indeed from this point of view we can interpret the one-dimensional Hilbert space as simply telling us that for quantum mechanics in a closed universe there can be no external observer.\footnote{We leave the theological implications of this statement as an exercise for the reader.}  For some previous comments along these lines see \cite{wheeler1980beyond,Hartle:2015vfa,Chandrasekaran:2022cip,Witten:2023qsv,Shaghoulian:2023odo}; our new contribution is a concrete mathematical formalism to make the idea precise, which allows us to quantitatively evade the apparent problems arising from a one-dimensional Hilbert space.  Some other recent work using observers in quantum gravity to address various problems is \cite{Lennytalk,Maldacena:2024spf,Jensen:2024dnl}.

\bfig
\includegraphics[height=5cm]{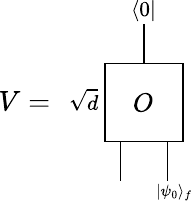}
\caption{Encoding the effective field theory Hilbert space of a close universe into a one-dimensional fundamental Hilbert space.}\label{Ocodefig}
\efig
We can state our proposal more concretely in terms of a linear encoding map $V:\mathcal{H}_{eff}\to\mathcal{H}_{fund}$, where $\mathcal{H}_{eff}$ is the Hilbert space of gravitational effective field theory and $\mathcal{H}_{fund}$ is the fundamental Hilbert space of the quantum gravity theory.  In AdS/CFT, $\mathcal{H}_{fund}$ is the Hilbert space of the dual CFT. $V$ often has the features of a quantum error correcting code \cite{Almheiri:2014lwa}, and in particular the quantum extremal surface formula can be viewed as a consequence of this structure \cite{Harlow:2016vwg}.  The basic ingredient of our framework for quantum gravity in a closed universe is the encoding map shown in figure \ref{Ocodefig}: we tensor our effective closed universe state with some fixed state $|\psi_0\ran_f$, act with some fixed orthogonal transformation $O$, and then project onto a definite pure state $\lan 0|$.  Written as an equation,
\be
V|\psi\ran=\sqrt{d}\lan 0|O\left(|\psi\ran\otimes|\psi_0\ran_f\right).
\ee
Here $|\psi\ran$ is a state in $\mathcal{H}_{eff}$ and $d$ is the dimensionality of $O$.  Naively $d$ is the dimensionality of $\mathcal{H}_{eff}$, but we include $|\psi_0\ran_f$ to give us the freedom to restrict which states we feed in (i.e. by lowering the cutoff on the gravitational effective field theory).  For our simplest code model, where we take $O$ to be generic in the Haar measure, we will see that we can interpret $1/d$ as being analogous to whatever parameter suppresses higher topologies in the gravitational path integral.\footnote{In a more refined version of the code, where we give $O$ a bit more internal structure, we will instead take $d\to \infty$, with the topological suppression coming from a bottleneck inside of $O$.}   $O$ is orthogonal (instead of just unitary) because in quantum gravity $\CRT$ must be a gauge symmetry, so in a closed universe there is a natural real structure for the theory \cite{Harlow:2023hjb}.  More concretely there is a basis of $\CRT$-invariant states in which $O$ is real, and more general $\CRT$-invariant states are superpositions of these with real coefficients. The factor of $\sqrt{d}$ is included so that this transformation on average (in the Haar measure on $O(d)$) preserves the inner product:
\be
\int dO \lan \phi|V^\dagger V|\psi\ran=\lan\phi|\psi\ran.
\ee
See appendix \ref{Oapp} for a review of how to compute such integrals.  On the other hand this map cannot possibly preserve the inner product for a fixed $O$, as it maps all of $\mathcal{H}_{eff}$ to a one-dimensional Hilbert space.  This is reflected in the typical size of fluctuations if we take $O$ to be a generic sample from the Haar measure on $O(d)$:\footnote{Here we introduce a convention where $\mO$ means ``of order'', to distinguish it from the orthogonal transformation $O$.}
\be
\int dO \Big|\lan \phi|V^\dagger V|\psi\ran-\lan \phi|\psi\ran\Big|^2=1+|\lan \phi^*|\psi\ran|^2+\mO\left(\frac{1}{d}\right).\label{Ofluct}
\ee
Here $|\phi^*\ran$ is the $\CRT$-conjugate of $|\phi\ran$; since $\CRT$ is a gauge symmetry we should require that $|\phi^*\ran=|\phi\ran$.  The essential point however is that the right-hand side of \eqref{Ofluct} is $\mO(1)$; the inner product is not close to being preserved by the map $V$.  In the next section we will see that this precisely matches what is seen from the gravitational path integral, so it is a feature rather than a bug, but we still need to learn how to make sensible predictions in this framework.  

\bfig
\includegraphics[height=6cm]{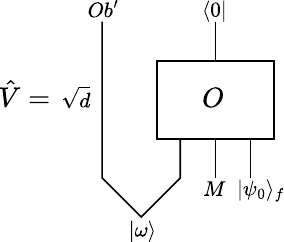}
\caption{Modified encoding map $\hat{V}$ including an entangled observer $Ob'$  who is cloned out of the system.  The orthogonal map $O$ is the same as in figure \ref{Ocodefig}.}\label{Ocode2fig}
\efig
Our proposal for computing observables is to introduce an external ``cloned observer'' system $Ob'$ which is entangled with some of the input to $O$ in a fixed state $|\omega\ran$, as shown in figure \ref{Ocode2fig}, after which we can re-interpret the map $V$ as a map $\hat{V}:\mathcal{H}_{M}\to \mathcal{H}_{Ob'}$. Roughly speaking $\mathcal{H}_{M}$ is the ``matter'' degrees of freedom which the observer would like to measure.  We explain the motivation for this rule in section \ref{Obsec}, but briefly the entanglement arises from the entanglement of the observer with the rest of the universe due to decoherence, and the observer is cloned out of the system in their pointer basis to ensure that they are effectively classical.  The real point however is that for the map $\hat{V}$, we will see that  the fluctuations \eqref{Ofluct} of the encoded inner product are suppressed by a factor of $e^{-S_2(\omega_{Ob'})}$, where $S_2$ is the second Renyi entropy and $\omega_{Ob'}$ is the reduction of $|\omega\ran$ to $Ob'$.  We will argue that $S_2(\omega_{Ob'})$ needs to be of order the coarse-grained entropy $S_{Ob}$ of the observer, so the inner product is preserved up to effects which are exponentially small in $S_{Ob}$.
For example if the observer is a human this suppression is of order $e^{-10^{24}}$. 
 Approximately preserving the inner product is a sufficient condition for constructing a (in general nonlinear) measurement theory in the fundamental description \cite{Akers:2022qdl}, as we will briefly review in the final discussion.

This claim may seem somewhat audacious.  Typically in quantum gravity we expect effective field theory to be valid up to effects which are of order $e^{-\frac{A}{G}}$, where $A$ is some characteristic area of the system.  For example in AdS near vacuum $A$ is set by the AdS scale, while in the black hole interior it is set by the horizon area (we should also restrict to observables whose complexity is not exponential in $A/G$). Usually such discussions do not require treatment of the observer as a concrete physical system with a finite entropy and energy.  Why are things different in a closed universe?  We think the basic reason is that in a closed universe there is a fundamental limit to how entropic an observer can be before their gravitational backreaction collapses the universe in a big crunch.  There is therefore a fundamental limit on the precision of observations in such a universe, and there is no reason for a theory to make predictions which are more precise than what any observer can observe \cite{Harlow:2010my,Banks:2002wr}.  In the AdS and black hole examples we can avoid this issue by putting our observer at the asymptotic boundary, and in this limit we can make the observer as heavy and long-lived as we like.  They therefore can record and process an arbitrary amount of information, so  there can be a precise formulation of the theory which is independent of the details of the observer.  On the other hand, even in AdS if the observer stays within the bulk gravitational system there are limitations on how big they can be; we can think of the order $e^{-\frac{A}{G}}$ errors mentioned above as arising from the entropy of the ``largest possible'' observer who can fit in the system without undergoing gravitational collapse.  Our goal for the rest of this paper is to explain how these somewhat vague philosophical statements are realized via the observer model shown in figure \ref{Ocode2fig}, and also to show how they are compatible with the results of the gravitational path integral.

There are several related works in the previous literature.  In \cite{Dong:2020uxp} the authors used tensor networks to show that information can be gotten out of a closed universe through entanglement with a reference system. This is essentially the same mechanism we use here, although the motivation and interpretation were different.  Similarly as this work was being completed \cite{Sahu:2024ccg} appeared, which points out that if we entangle a black hole in a closed universe with some external system then we can view this external system as providing a nontrivial Hilbert space for the closed universe.  Mathematically this is again the same setup as shown in figure \ref{Ocode2fig}, so their mechanism for ``growing'' a Hilbert space is the same as ours.  There were also some papers in which a closed universe is in the island of some asymptotically AdS spacetime dual to some holographic CFT's \cite{Antonini:2022blk,Antonini:2023hdh,Antonini:2024mci}. If we identify our external cloned observer $Ob'$ with their holographic CFT, the mathematics is again similar.  Our interpretation of this mathematics is quite different however, as we have a single observer instead of two black holes or holographic CFT's and for us the entanglement is a necessary consequence of interaction between the observer and its environment together with the cloning of the observer out of the system rather than a choice of initial state of a closed universe together with something else. In \cite{Balasubramanian:2023xyd} the authors entangled de Sitter space with an AdS black hole and considered the black hole as a kind of observer. Our setup differs from this one both at the level of interpretation but also mathematically; for us the external cloned observer $Ob'$ is non-gravitational and the entropy of the closed universe behaves smoothly as a function of the observer's entropy.

Our plan for the rest of this paper is the following: in section \ref{revsec} we review several of the arguments for a one-dimensional Hilbert space and explore some of their consequences.  In section \ref{Obsec} we review the inevitability of entanglement between an observer and its environment in a pointer basis and use this to motivate our construction.  In section \ref{closedobsec} we explain how including the observer in a closed universe leads to a nontrivial Hilbert space whose dimension is upper bounded by $S_{Ob}$.  In section \ref{bhobsec} we show that including an observer inside of a black hole gives a Hilbert space whose dimension is upper bounded by $S_{Ob}+S_{BH}$.  In section \ref{discussionsec} we discuss some remaining high level issues.  Technical results are included in a set of appendices.

\section{Why is the Hilbert space one-dimensional?}\label{revsec}
In this section we'll review several of the arguments for a one-dimensional closed universe Hilbert space, and then show how this (unsurprisingly) leads to problems with applying the rules of quantum mechanics in a closed universe in the standard way.  

\subsection{Arguments for a one-dimensional Hilbert space}
We give three main arguments that the Hilbert space of a closed universe is one-dimensional:

\bfig
\includegraphics[height=5cm]{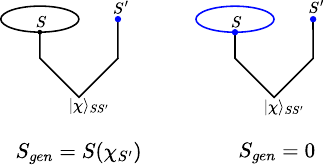}
\caption{Competing entanglement wedges, shaded in blue, for a reference system $S'$ entangled with a system $S$ in a closed universe.  On the left is the case where the entanglement wedge doesn't include the closed universe, while on the right is the case where it does.  The wedge on the right always wins, so it is impossible to be nontrivially entangled with a closed universe.}\label{referencewedgefig}
\efig
\textbf{Entanglement wedge of a reference system:} Consider a quantum system $S$ in a closed universe that in the effective description is entangled with some reference system $S'$ in a state $|\chi\ran_{SS'}$, as in figure \ref{referencewedgefig}.  We can use the quantum extremal surface formula \cite{Engelhardt:2014gca} to compute the entropy of $S'$ in the fundamental description of the system.  There are two candidates for the entanglement wedge of $S'$: $S'$ itself, or the $S'$ plus the entire closed universe.  According to the quantum extremal surface formula the entropy of $S'$ in the fundamental description is therefore given by
\be
S(\rho_{S'})=\min \Big(S(\chi_{S'}),0\Big)=0,
\ee
where $\chi_{S'}$ denotes the reduction of the state $|\chi\ran_{SS'}$ to the subsystem $S'$, so in the fundamental description $S'$ is in a pure quantum state regardless of our choice of $|\chi\ran_{SS'}$ \cite{Almheiri:2019hni}.  In other words the holographic encoding of $S$ into the fundamental description should be proportional to a rank one projection to ensure that $S'$ is always pure.

\bfig
\includegraphics[height=5cm]{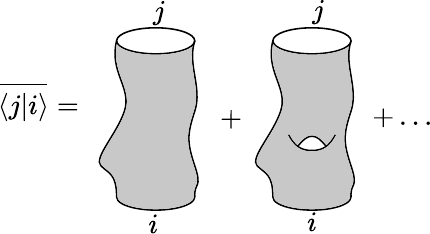}
\caption{Using the gravitational path integral to compute the average of the closed universe inner product.}\label{ipfig}
\efig  
\textbf{Rank of the Gram matrix:} Consider a set of states $|i\ran$ for a closed universe.  We can define the inner product matrix
\be\label{naiveip}
M_{ji}=\lan j|i\ran,
\ee 
also called the Gram matrix, and it is straightforward to argue that the Hilbert space spanned by the $|i\ran$ is one-dimensional if and only if $M$ obeys
\be\label{Mreq}
\Tr\left(M\right)^n=\Tr\left( M^n\right)
\ee
for $n$ any positive integer.  Naively we can compute $M$ using the gravitational path integral, see figure \ref{ipfig}, but we need to be careful about the fact that the path integral is really computing some kind of coarse-grained average rather than literally giving us the components of $M_{ji}$ in a fixed theory \cite{Saad:2019lba,Penington:2019kki}.   So the path integral shown in figure \ref{ipfig} is really computing $\ol{M_{ji}}=\ol{\lan j|i\ran}$, where $|i\ran$ and $|j\ran$ are states in the Hilbert space of some fundamental quantum gravity theory whose low-energy effective action is the one appearing in the path integral and the bar indicates the coarse-graining average.   Which states they are is determined by the boundary conditions of the path integral in the future and past.  To test \eqref{Mreq} using the path integral we therefore need to directly compute averages of the left and right hand sides of \eqref{Mreq}.
\bfig
\includegraphics[height=6cm]{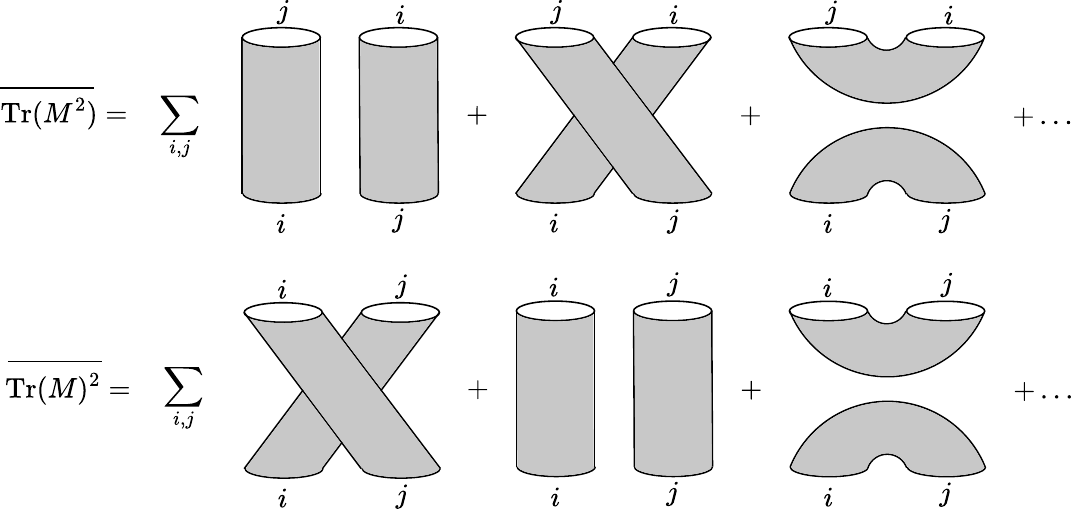}
\caption{Computing the averages of the two sides of \eqref{Mreq} using the gravitational path integral; they are equal by the permutation invariance of the set of things we sum over \cite{Penington:2019kki}.}\label{gram2fig}
\efig
This is shown for $n=2$ in figure \ref{gram2fig}, where we see that, due to the permutation invariance of the geometries which are summed over, the two averages are equal.  This statement holds for all $n$, so \eqref{Mreq} holds at least on average \cite{Penington:2019kki}.  In fact the same permutation invariance also shows 
\be
\ol{\left(\Tr\left( M\right)^n-\Tr\left( M^n\right)\right)^2}=0,
\ee
so there are no fluctuations in \eqref{Mreq} and the Hilbert space spanned by the $|i\ran$ is one-dimensional for each instance in the ensemble over which we are averaging.  This is true whatever set of states we choose, so the full Hilbert space is one-dimensional.

\bfig
\includegraphics[height=4cm]{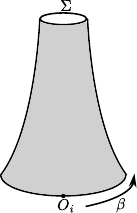}
\caption{Preparing a closed universe state $|i\ran$ on a Cauchy slice $\Sigma$ using the gravitational path integral with a past Euclidean AdS boundary of length $\beta$ with an operator $O_i$ inserted.}\label{stateprepfig}
\efig   
\bfig
\includegraphics[height=4cm]{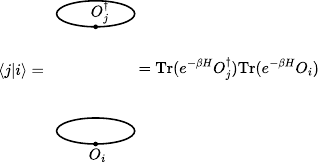}
\caption{Using the dual CFT to compute the inner product between two closed universe states.}\label{closedipfig}
\efig
\textbf{Inner product in a concrete CFT dual:} The previous argument used the path integral, and thus had to deal with some kind of average.  Here is an argument that works directly in a fixed theory \cite{Usatyuk:2024mzs}.  Consider a holographic CFT, say $\mathcal{N}=4$ super Yang Mills theory.  In the dual bulk theory there are crunching closed universe solutions, and if we continue to Euclidean signature we can prepare states of this type by using a complete AdS boundary of length $\beta$ in past Euclidean time with a boundary operator $O_i$ inserted (see figure \ref{stateprepfig}) \cite{Maldacena:2004rf,McInnes:2004ci,VanRaamsdonk:2020tlr}. If we use the gravitational path integral to compute the inner product of these states we are led back to figure \ref{ipfig}, but we can instead use the dual CFT as shown in figure \ref{closedipfig}.  This shows that the inner product is
\be
\lan j|i\ran=\Tr\left(e^{-\beta H}O_j^\dagger\right) \Tr\left(e^{-\beta H}O_i\right),
\ee
which manifestly has rank one.  So again we see the Hilbert space is one-dimensional.

Another argument for a one-dimensional Hilbert space that is worth mentioning, based on the idea that there should be no ``$-1$ form global symmetries'' in quantum gravity, was given \cite{McNamara:2020uza}.\footnote{The basic idea of a $-1$-form global symmetry is that any local operator $O$ can be viewed as a ``conserved current'' for a one-form symmetry, since $\star O$ is a top form that necessarily obeys $d\star O=0$.  We can therefore turn on a ``background gauge field'' for the symmetry by adding a term $\lambda O$ to the Lagrangian where the coupling constant $\lambda$ is the gauge field.  Saying there are no one-form global symmetries is therefore the same as saying there are no adjustable coupling constants in the theory. 
 The argument of \cite{McNamara:2020uza} is that a nontrivial baby universe Hilbert space would lead to an average over coupling constants as advocated by Coleman \cite{Coleman:1988cy}, which would violate the conjecture that there are no adjustable coupling constants.}  See also \cite{Gesteau:2020wrk} for a more mathematical perspective. 

\subsection{Consequences of a one-dimensional Hilbert space}
\bfig
\includegraphics[height=3.5cm]{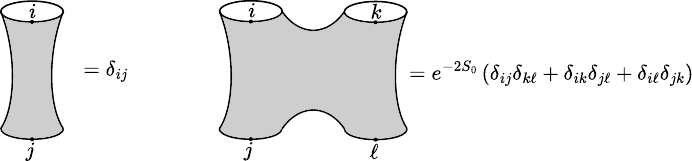}
\caption{Sample calculations in the topological model with matter.}\label{topexfig}
\efig
We'll now use the gravitational path integral to illustrate various pathologies which arise from the one-dimensional nature of the Hilbert space.  We will see that these pathologies can all be interpreted in terms of a holographic map of the type shown in figure \ref{Ocodefig}. To be concrete we will use a topological model of quantum gravity in $1+1$ dimensions that was introduced in \cite{Usatyuk:2024mzs}, which generalizes the topological model of \cite{Marolf:2020xie} to include a simple kind of topological matter.  As in \cite{Marolf:2020xie}, the gravitational action on a Euclidean spacetime manifold $\mathcal{M}$ is given by
\be
I_E[\mathcal{M}]=-S_0\chi[\mathcal{M}],
\ee
where
\be
\chi[\mathcal{M}]=\frac{1}{4\pi}\int_\mathcal{M} R+\frac{1}{2\pi}\int_{\partial \mathcal{M}}K=2-2g-b
\ee
is the Euler character ($g$ is the genus and $b$ is the number of boundaries).  The matter is described by a worldline which carries an index $i$.  These worldlines are created at points on the asymptotic boundary, and their bulk dynamics is simply a rule that each boundary index must pair up with another boundary index with the same value that is part of the same connected component in the bulk.  In other words the ``propagator'' for each worldline index is just $\delta_{ij}$.  We can think of this as modeling the worldline action of a heavy particle in the gravitational system.\footnote{In this model we do not include matter loops or keep track of homotopically distinct paths between the same boundary endpoints; we could include these, but then we'd need to introduce another parameter analogous to $S_0$ to suppress more complicated paths.  In the case of heavy matter this suppression arises automatically from the worldline action.}  
See figure \ref{topexfig} for some examples of calculations in this model. We can view the boundaries as creating and annihilating states of the closed universe, so for example the first diagram in figure \ref{topexfig} is a contribution to the averaged inner product between a closed universe state $|i\ran$ where the particle is in state $i$ and a closed universe state $|j\ran$ where the particle is in state $j$.  If we include all the higher genus contributions to this inner product we get
\be
\ol{\lan j|i\ran}=\frac{\delta_{ij}}{1-e^{-2S_0}},
\ee
so as defined the states aren't quite normalized to one.  

\bfig
\includegraphics[height=3.8cm]{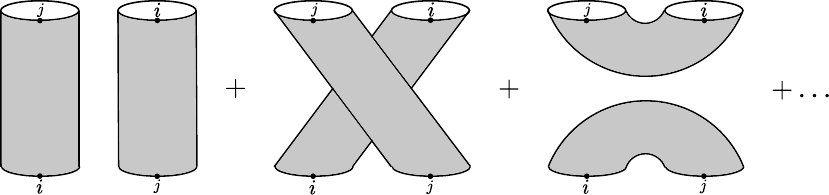}
\caption{Computing the square of the inner product in the topological model.}\label{topip2fig}
\efig
To see the imprint of the one-dimensional Hilbert space, we can look at the average of the square of this inner product \cite{Usatyuk:2024mzs}.  The leading contributions are shown in figure \ref{topip2fig}, they evaluate to
\be\label{ip2top}
\ol{\lan j|i\ran\lan i|j\ran}=\delta_{ij}+1+\delta_{ij}+\mO\left(e^{-2S_0}\right).
\ee
We can use this to also study the fluctuations of the inner product:
\be
\ol{|\lan j|i\ran-\ol{\lan j|i\ran}|^2}=\ol{\lan j|i\ran\lan i|j\ran}-|\ol{\lan j|i\ran}|^2=1+\delta_{ij}+\mO\left(e^{-2S_0}\right).
\ee
Comparing this expression to \eqref{Ofluct} we see an exact match: the inner product has large fluctuations, with the two terms on the right-hand side of \eqref{Ofluct} coming in the gravitational calculation from the second and third saddles in figure \ref{topip2fig}.  Corrections are suppressed by $e^{-2S_0}$, which we interpret as $1/d$ in the code model.  Note that for this match to work it is important that we are interpreting the $|i\ran,|j\ran,\ldots$ states we are averaging over as encoded states in the fundamental description.  Also we emphasize that had we integrated over unitary matrices instead of orthogonal matrices in \eqref{Ofluct}, we wouldn't have gotten the term corresponding to the third saddle in figure \ref{topip2fig}; this is one illustration of the automatic gauging of $\CRT$ by the Euclidean gravity path integral \cite{Harlow:2023hjb}.

Another way to phrase this problem is in terms of an observable: if we start the system in the state $|j\ran$ and then measure the observable $|i\ran\lan i|$, the expectation value is given by \eqref{ip2top}.  The semiclassical result from canonical gravity would just be $\delta_{ij}$, so the two additional terms are large corrections to the canonical result.

\bfig
\includegraphics[height=3.7cm]{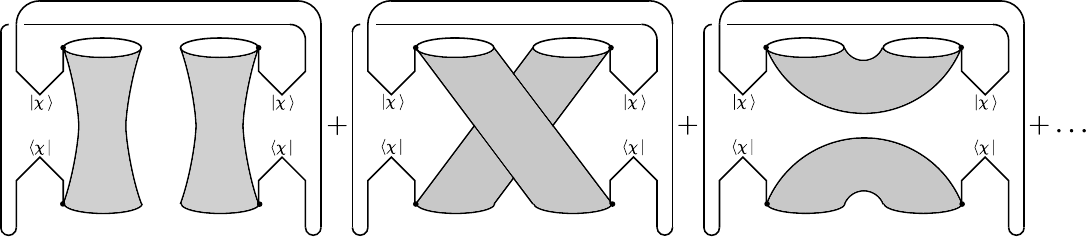}
\caption{Computing the second Renyi entropy of a purifying matter system.}\label{renyitopfig}
\efig
Yet a third way to capture this failure to preserve the inner product is by entangling the wordline index $i$, which in the effective description we can think of as spanning a quantum matter system $M$, with a reference system $M'$ in a state $|\chi\ran_{MM'}$, and then using the path integral to compute the average of the second Renyi entropy of the reference system in the fundamental description.  The leading contributions to this calculation are shown in figure \ref{renyitopfig}, the result is
\be\label{toprenyi}
\ol{e^{-S_2(\Psi_{M'})}}=e^{-S_2(\chi_{M'})}+1+\Tr\left(\chi_{M}\chi_{M}^T\right)+\mO(e^{-2S_0})
\ee
where $\chi_{M'}$ is the reduction of $|\chi\ran$ and $\Psi_{M'}$ is the reduction to $M'$ of a state $|\Psi\ran$ in the fundamental description whose averaged Renyi entropy we interpret the path integral as computing.
In general we have $0\leq \Tr\left(\chi_{M}\chi_{M}^T\right)\leq e^{-S_2(\chi_{M'})}$, and if we take $|\chi\ran$ to be $\CRT$-invariant then $\chi_{M}^T=\chi_{M}$ so $\Tr\left(\chi_{M'}\chi_{M'}^T\right)=e^{-S_2(\chi_{M})}=e^{-S_2(\chi_{M'})}$. Note in particular that the right-hand side is never small, so the entropy of the reference system is never large. 
 We can compare this result to a code calculation using our encoding map from figure \ref{Ocodefig} with $O$ taken at random in the Haar measure: defining
\be
|\Psi\ran=(V\otimes I_{M'})|\chi\ran_{MM'},
\ee
we indeed have
\be
\int d O \Tr\left(\Psi_{M'}^2\right)=e^{-S_2(\chi_{M'})}+1+\Tr\left(\chi_{M}\chi_{M}^T\right)+\mO\left(\frac{1}{d}\right).\label{coderenyi1}
\ee
The code makes the distinction between the states $\Psi_{M'}$ and $\chi_{M'}$ on the left and right hand sides of \eqref{toprenyi} quite clear: $|\chi\ran$ is a state living in the effective description, but what the path integral is really computing is averages of inner products and entropies for the encoded state $|\Psi\ran$ in the fundamental description.

We've occasionally encountered the viewpoint that perhaps the way to avoid the problem of a one-dimensional Hilbert space is to consider a larger Hilbert space spanned by the unique states for each of the theories the path integral is averaging over.  From this point of view these states are sometimes called ``$\alpha$-states''.  This however does not help with the problem of large fluctuations of observables, as is clear from equation \eqref{ip2top}.  In any event in this paper our perspective is that we are always really working in a fixed theory, which from the coding point of view is defined by the choice of the orthogonal matrix $O$ in figure \ref{Ocodefig}, and so we do not have distinct $\alpha$-states in the theory.  The averages we perform over $O$ are purely to show us what the behavior will be like in a typical fixed $O$.

\section{Entanglement and observers}\label{Obsec}
We'd now like to include an observer in the story.  We obviously prefer to avoid any detailed discussion of what constitutes an observer, but we need to make three basic assumptions:
\begin{enumerate}
\item An observer has a Hilbert space dimension $e^{S_{Ob}}$ whose size controls the precision of the experiments the observer can do.  More concretely no observer can make measurements to accuracy $e^{-c S_{Ob}}$ with $c>0$ in the limit that $S_{Ob}\to \infty$.
\item The observer is \textit{classical}, in the sense that it has a basis of ``pointer'' states $|a\ran_{Ob}$ which are stable under interaction with its environment $E$. 
\item The observer is indeed entangled with its environment, with a density matrix in the pointer basis that is close to diagonal and an entanglement entropy of order $S_{Ob}$.
\end{enumerate}
To motivate these assumptions, recall that a pointer state $|\psi\ran_S$ of a system $S$ interacting with an environment $E$ is one where
\be
\Tr_E\left(U_{SE}\left(|\psi\ran\lan \psi|_S\otimes \frac{I_E}{|E|}\right)U_{SE}^\dagger\right)\approx |\psi\ran\lan \psi|_S
\ee
where $U_{SE}$ is the time evolution operator in the interaction picture.  A density matrix which is diagonal in a basis of pointer states will be preserved under interaction with the environment, while off-diagonal elements will be suppressed via the mechanism of decoherence \cite{Zurek:2003zz}.  It is not obvious that pointer states exist.  To understand why they might, we can write the interaction Hamiltonian as
\be
H_{SE}=\sum_p O_S^p \otimes O_E^p,
\ee
where $O_S^p$ and $O_E^p$ are bases for the operators on $S$ and $E$ respectively.  In particular if there is one term $p_0$ that dominates the sum, then the pointer basis is just the eigenstates of $O_S^{p_0}$.  If there are multiple competing terms whose $O_S^p$ commute then we we can simultaneously diagonalize them.  If there are multiple competing terms whose $O_S^p$ don't commute, then we should use coherent states.  

A simple example to keep in mind is the situation where $S$ and $E$ each consist of one qubit and $U_{SE}$ is just the entangling CNOT gate 
\be
U_{SE}|a,b\ran=|a,b+a\ran,
\ee
with the addition being done $\mathrm{mod}\, 2$.  The pointer basis for the first qubit is just the z-basis, and indeed we have
\begin{align}
\Tr_E\left(U_{SE}\left(|a\ran\lan b|\otimes \frac{I}{2}\right)U_{SE}^\dagger\right)&=\delta_{ab}|a\ran\lan b|
\end{align}
so off-diagonal components of the density matrix in this basis are indeed removed by the evolution.  Viewed as a transformation of states on $S$ this is an example of what is called a \textit{quantum-to-classical channel} \cite{Watrous:2018zil}, which is a completely positive trace-preserving map $\mathcal{C}$ such that in some basis we have
\be
\mathcal{C}(|a\ran\lan b|)=\delta_{ab}|a\ran\lan b|.  
\ee
In this example the decoherence may appear fine-tuned, but for systems with a more complicated environment it happens generically on a timescale which is exponentially small in the size of the environment.  This choice of a preferred basis by interaction with the environment is sometimes called \textit{einselection}, and it is the way an observer can have a stable classical experience of the world despite constant interaction with a quantum environment \cite{Zurek:2003zz}.  Indeed this interaction is unavoidable, so an observer is always substantially entangled with the rest of the world in their pointer basis.  If we now evolve the system in the Schrodinger picture (including the internal dynamics of the system and the environment in addition to the interaction Hamiltonian), decoherence ensures that the density matrix of the system remains diagonal in the pointer basis so the only evolution is that of the diagonal components in this basis.  We can view this as a time-dependent classical probability distribution.

Returning now to our closed universe setting, we need to account for the entanglement of the observer with their environment in a pointer basis.  Indeed even if there is no obvious environment, at a minimum the observer should at least be entangled with their gravitational field since a reasonable observer must be somewhat localized in time and this requires them to be in a broad superposition of energy eigenstates.  These must then be dressed by the gravitational field, roughly as 
\be
|\psi\ran=\sum_aC_a|E_a\ran_{Ob}|g_a\ran_{Gr},
\ee
and the range of eigenstates contributing nontrivially to this sum should be $\mO\left(e^{S_{Ob}}\right)$.  For recent progress on gravitational entanglement see \cite{Danielson:2021egj,Danielson:2022tdw,Biggs:2024dgp}.  This will not be the dominant form of entanglement in situations with a richer environment however, since the matter in such an environment presumably interacts more strongly with the observer than gravity does, but in any event we will not be so concerned about the details of the entanglement between the observer and its environment - it is enough for us that it is there.  

\bfig
\includegraphics[height=6cm]{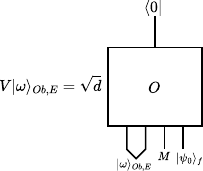}
\caption{An observer $Ob$ entangled with an environment $E$ and some matter $M$ being fed into the holographic map $V$ for a closed universe.}\label{obsinfig}
\efig
So far we've motivated a code picture as in figure \ref{obsinfig}, where an observer $Ob$ entangled with some environment $E$, together with some additional degrees of freedom we'll call $M$ (which we can loosely think of as the ``matter'' on which the observer will do experiments), are encoded by a holographic map $V$ into the unique state of a closed universe.\footnote{There of course are also entangling interactions between $E$ and $M$, we think of these as being included in $O$.}  This has the problem however that the observer is treated on the same footing as the matter, and in particular the observer can be in a non-classical state where their density matrix is not close to diagonal in the pointer basis.  It also has the problem of the one-dimensional Hilbert space, which for example implies that the inner product of two states of this type factorizes as shown in the left side of figure \ref{alternatefig}.  Our central point is that we can fix both of these problems by modifying this rule for the inner product, which we do by passing $Ob$ through the quantum-to-classical channel $\mathcal{C}_{Ob}$ in their pointer basis.  See figure \ref{alternatefig} for an illustration.  This ensures that the state of the observer is classical, and the inner product also no longer factorizes so there is a chance for a more nontrivial quantum theory.
\bfig
\includegraphics[height=8cm]{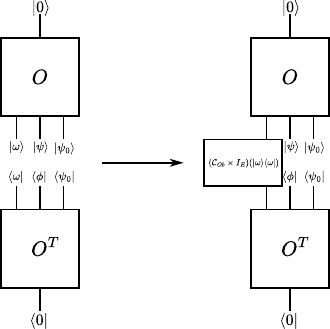}
\caption{Modifying the rule for the inner product by applying a quantum-to-classical channel to the observer, disrupting the product structure of the inner product.}\label{alternatefig}
\efig 

\bfig
\includegraphics[height=6cm]{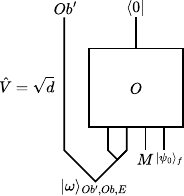}
\caption{Cloning the observer out of the system.  Before acting with $\lan 0|O$ the cloned observer $Ob'$, the observer $Ob$, and the environment $E$ are in a shared GHZ-type state in the pointer basis.  We can interpret this as defining a new holographic map $\hat{V}:M\to Ob'$.}\label{obsoutfig}
\efig
There is an alternative way to visualize this rule which makes the nature of the holographic map more transparent.  This is that we can view the quantum-to-classical channel $\mathcal{C}_{Ob}$ as resulting from a unitary ``cloning'' operation,\footnote{General cloning of quantum states is of course impossible \cite{Wootters:1982zz}, but there is no obstruction to cloning in a particular basis and here we have a particular basis at hand: the pointer basis $|a\ran_{Ob}$.} which acts on $Ob$ together with a ``clone'' $Ob'$ as 
\be
|0\ran_{Ob'}|a\ran_{Ob}\to  |a\ran_{Ob'}|a\ran_{Ob}.
\ee
Tracing out $Ob'$ gives back our quantum-to-classical channel, but it is more useful to keep $Ob'$ around since we can then re-interpret this definition as introducing a new holographic map $\hat{V}:\mathcal{H}_M\to \mathcal{H}_{Ob'}$ as shown in figure \ref{obsoutfig}.  Because this cloning happens in the pointer basis, from the point of view of the original observer $Ob$ their clone $Ob'$ is just another part of the environment, and in particular this cloning will not disrupt their classical experience.  We can use $\hat{V}$ to construct a nontrivial Hilbert space for $Ob$ to use when they make predictions of what they will see making measurements on $M$.\footnote{To match figure \ref{obsoutfig} to figure \ref{Ocode2fig} we can simply merge the $E$ and $Ob$ lines going into $O$, as this simplifies calculations without changing any of the results.}    In the rest of the paper we will see that this leads to sensible results.

\section{An observer in a closed universe}\label{closedobsec}
We now revisit the arguments of section \ref{revsec} for the encoding map $\hat{V}$ that includes the external cloned observer $Ob'$ entangled with the original observer $Ob$ and environment $E$ in the closed universe.  We will see that all of the problems arising from the one-dimensional Hilbert space are now exponentially suppressed in $S_{Ob}$.  We will do calculations both using the holographic code in figure \ref{Ocode2fig} and the gravitational path integral, with the latter being evaluated first for the topological model of \cite{Marolf:2020xie,Usatyuk:2024mzs} and then for JT gravity \cite{Jackiw:1984je,Teitelboim:1983ux}.  We will also explain how similar results can be obtained in the special case of a closed universe with negative cosmological constant using the eigenstate thermalization hypothesis to do calculations in the dual CFT \cite{Srednicki:1994mfb}.

\subsection{A simple code model}
\label{sec:code_generic}
We first discuss some calculations in  $\hat{V}:\mathcal{H}_M\to \mathcal{H}_{Ob'}$ code in the case where we take $O$ to be a generic sample from the Haar measure on $O(d)$.  For convenience we absorb $E$ from figure \ref{obsoutfig} into $Ob$, to get the code shown in figure \ref{Ocode2fig}.  All calculations use the orthogonal integration technology described in appendix \ref{Oapp}, here we will just quote results.  We first note that on average this code preserves the inner product on $\mathcal{H}_M$:
\be
\int dO \lan \phi|\hat{V}^\dagger \hat{V}|\psi\ran=\lan \phi|\psi\ran.
\ee
This was also true for the $V$ code from figure \ref{Ocodefig} however, so to get a better sense of what is going on for a typical fixed $O$ we should look at the fluctuations.  These are given by
\begin{align}\nonumber
&\int dO|\lan \phi|\hat{V}^\dagger \hat{V}|\psi\ran-\lan \phi|\psi\ran|^2\\
=&\ \frac{d}{d+2}\Bigg[|\lan\phi|\psi\ran|^2+e^{-S_2(\omega_{Ob'})}+\Tr\left(\omega_{Ob}\omega_{Ob}^T\right)|\lan\phi^*|\psi\ran|^2\Bigg]-|\lan\phi|\psi\ran|^2\nonumber\\
=&\ e^{-S_2(\omega_{Ob'})}+\Tr\left(\omega_{Ob}\omega_{Ob}^T\right)|\lan\phi^*|\psi\ran|^2+\mO\left(\frac{1}{d}\right),\label{codeipobsresult}
\end{align}
see appendix \ref{Oapp} for details of the calculation.  In general we have $0\leq \Tr\left(\omega_{Ob}\omega_{Ob}^T\right)\leq e^{-S_2(\omega_{Ob})}=e^{-S_2(\omega_{Ob'})}$, and taking $|\omega\ran\otimes |\psi\ran$ to be $\CRT$-invariant (i.e. real) we simply have
\be
\int dO|\lan \phi|\hat{V}^\dagger \hat{V}|\psi\ran-\lan \phi|\psi\ran|^2=(1+|\lan\phi|\psi\ran|^2)e^{-S_2(\omega_{Ob'})}+\mO\left(\frac{1}{d}\right).
\ee
In other words the fluctuations of the inner product are exponentially suppressed in $S_2(\omega_{Ob'})$, so a more complicated observer gets a more precise theory just as promised in the introduction.  

We can also phrase this in terms of the encoded Renyi entropy of a reference system $M'$ entangled with the matter system $M$, as we did without the observer in equation \eqref{coderenyi1}.  Now including the observer we can define
\be
|\Psi\ran_{Ob',M'}=\left(\hat{V}\otimes I_{M'}\right)|\chi\ran_{MM'},
\ee
in terms of which we have
\begin{align}
\int dO e^{-S_2(\Psi_{M'})}=\frac{d}{d+2}\Bigg[e^{-S_2(\chi_{M'})}+e^{-S_2(\omega_{Ob'})}+\Tr\left(\chi_{M}\chi_{M}^T\right)\Tr\left(\omega_{Ob}\omega_{Ob}^T\right)\Bigg].\label{codeobrenyi}
\end{align}
Assuming both Renyi entropies are large compared to one  and working at large $d$, we can simplify this to
\be
\label{eq:entropy_code_generic}
S_2(\Psi_{M'})=\min\Big(S_2(\chi_{M'}),S_2(\omega_{Ob'})\Big).
\ee
In other words we can entangle up to $S_2(\omega_{Ob'})\sim S_{Ob}$ degrees of freedom with the closed universe: by including the observer we have grown a Hilbert space!  We now turn to recovering these results in gravity.

\subsection{Topological model}
\bfig
\includegraphics[height=5.5cm]{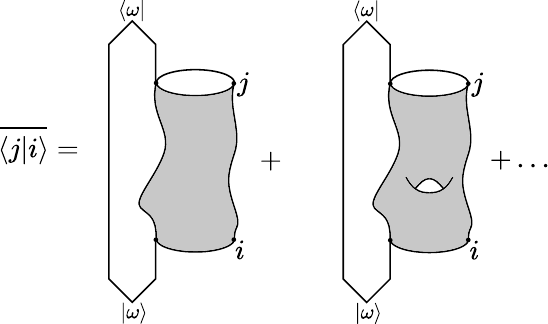}
\caption{Computing the closed universe inner product for the topological model in the presence of an observer.}\label{ipobsfig}
\efig
We can include an observer in the topological model of \cite{Marolf:2020xie,Usatyuk:2024mzs} by treating the internal state of the observer $Ob$ (and its environment $E$) as an additional type of worldline degree of freedom, this time with state label $a$ to distinguish it from the $i$ label for the matter world line.  As a first calculation we can compute the average of the inner product of states with different matter label, as in figure \ref{ipobsfig}.  Including the sum over genus we get
\be\label{topnorm2}
\ol{\lan j|i\ran}=\frac{\delta_{ij}}{1-e^{-2S_0}}
\ee
just as we did for the model with no matter. 

\bfig
\includegraphics[height=8cm]{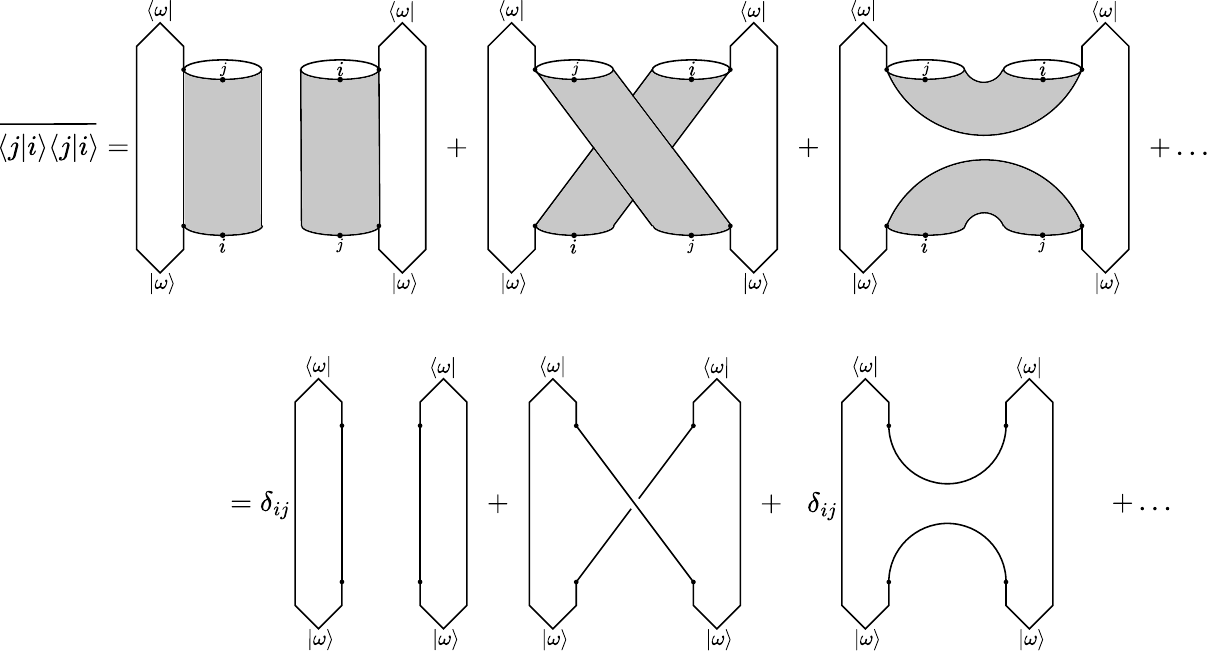}
\caption{Leading contributions to the square of the inner product in the topological model.}\label{topip2obsfig}
\efig
More interesting is the square of the inner product, the leading contributions to which are shown in figure \ref{topip2obsfig}.  Evaluating these contributions gives
\be
\ol{\lan j|i\ran\lan i|j\ran}=\delta_{ij}+e^{-S_2(\omega_{Ob'})}+\delta_{ij}\Tr\left(\omega_{Ob}\omega_{Ob}^T\right)+\mO(e^{-2S_0}),
\ee
so the inner product fluctuation is
\be \label{eqn:inner_prod_fluc_topmodel}
\ol{|\lan j|i\ran-\ol{\lan j|i\ran}|^2}=\ol{\lan j|i\ran\lan i|j\ran}-|\ol{\lan j|i\ran}|^2=e^{-S_2(\omega_{Ob'})}+\delta_{ij}\Tr\left(\omega_{Ob}\omega_{Ob}^T\right)+\mO(e^{-2S_0})
\ee
just as we found from a code calculation in \eqref{codeipobsresult}.  We can also entangle a reference system $M'$ with the worldline in an entangled state $|\chi\ran$ and use the path integral to compute the average of the second Renyi entropy.  The leading contributions are as in figure \ref{renyitopfig}, except that now there is also the observer entanglement as in figure \ref{topip2obsfig}.  The result is that
\be
\label{eq:Renyi_topological}
\ol{e^{-S_2(\Psi_{M'})}}=e^{-S_2(\chi_{M'})}+e^{-S_2(\omega_{Ob'})}+\Tr\left(\chi_{M}\chi_{M}^T\right)\Tr\left(\omega_{Ob}\omega_{Ob}^T\right)+\mO\left(e^{-2S_0}\right),
\ee
which again matches the leading order code result \eqref{codeobrenyi}.

We can also compute all-orders results in this model. The full set of topologies to sum over in figure \ref{topip2obsfig} consists of pairs of cylinders decorated by handles as well as connected geometries consisting of a sphere with four holes and some number of handles.  For the average inner product squared the full sum is
\begin{align}\nonumber
\ol{\lan j|i\ran\lan i |j\ran}=\frac{1}{(1-e^{-2S_0})^2}\Bigg[&\delta_{ij}+e^{-S_2(\omega_{Ob'})}+\delta_{ij}\Tr\left(\omega_{Ob}\omega_{Ob}^T\right)\\
&+e^{-2S_0}(1+e^{-2S_0})\left(1+2\delta_{ij}\right)\left(1+e^{-S_2(\omega_{Ob'})}+\Tr\left(\omega_{Ob}\omega_{Ob}^T\right)\right)\Bigg],\label{norm2topex}
\end{align}
where the second line is the contribution from connected geometries.  This contribution is strictly subleading if we assume that the observer entropy is small compared to $S_0$, which is the situation we have so far been considering.  If we do allow $S_{Ob}$ to be comparable to (or larger than) $S_0$, then a random choice of $O$ in the code no longer gives a good match for gravity.  The issue is that the connected contribution to the gravitational calculation has a term of order $e^{-2S_0}$ which is not multiplied by $\delta_{ij}$ or suppressed by the observer entropy, while there is no such term in the exact code result quoted in the first line of \eqref{codeipobsresult}.  This connected contribution can even beat the contributions from the crossed cylinders if the observer entropy is large compared to $S_0$.  That situation cannot be realized in the code model given our identification of $d\sim e^{S_0}$, since the observer isn't a valid input unless $e^{S_{Ob}}\leq d$.  In appendix \ref{app:structuredcode} we present a more structured code, where $O$ is not chosen completely at random, in which $e^{S_0}$ appears from an additional bottleneck in the code rather than the dimensionality of $O$, and we show that this code reproduces the connected contribution to $\ol{\lan j|i\ran\lan i|j\ran}$ from the gravitational result. 

There is an important subtlety in the expression \eqref{norm2topex} however, which is that the norms of the $|i\ran$ states are fluctuating random variables.  To really capture the fluctuations of the squared inner product at higher orders in $e^{-2S_0}$, we should instead compute the quantity 
$$\ol{\left(\frac{\lan j|i\ran\lan i|j\ran}{\lan i|i\ran\lan j|j\ran}\right)}.$$
The only way we know how to evaluate this using the path integral is to first compute $\ol{\lan j|i\ran\lan i|j\ran\big(\lan i|i\ran\lan j|j\ran\big)^m}$ and then analytically continue to $m=-1$, which is somewhat unpleasant.  This is overkill for our purposes however, as we are only interested in the situation where $S_0$ and $S_{Ob}$ are both large and in this situation the norm fluctuations (which we can obtain by setting $i=j$ in \eqref{norm2topex}) do not affect the leading expression
\be
\ol{|\lan j|i\ran-\ol{\lan j|i\ran}|^2}=e^{-S_2(\omega_{Ob'})}+\delta_{ij}\Tr\left(\omega_{Ob}\omega_{Ob}^T\right)+e^{-2S_0}+\ldots.\label{topips0}
\ee
Here $\ldots$ indicates terms that are suppressed by higher powers of $e^{-2S_0}$ and/or $e^{-S_{Ob}}$.  

Similar comments apply to the second Renyi entropy of $M'$.  We first have 
\begin{align}\nonumber
\ol{\Tr(\Psi_{M'}^2)}=\frac{1}{(1-e^{-2S_0})^2}\Bigg[&e^{-S_2(\chi_{M'})}+e^{-S_2(\omega_{Ob'})}+\Tr\left(\chi_{M}\chi_{M}^T\right)\Tr\left(\omega_{Ob}\omega_{Ob}^T\right)\\\nonumber
&+e^{-2S_0}(1+e^{-2S_0})\Big(1+e^{-S_2(\chi_{M'})}+\Tr\left(\chi_{M}\chi_{M}^T\right)\Big)\\
&\phantom{+}\times\Big(1+e^{-S_2(\omega_{Ob'})}+\Tr\left(\omega_{Ob}\omega_{Ob}^T\right)\Big)\Bigg],
\label{eq:entropy_topological}
\end{align}
where $|\Psi\ran$ is obtained by feeding properly normalized states $|\omega\ran$ and $|\chi\ran$ directly into the path integral without further adjustment of the normalization.  This is not really a computation of the Renyi entropy however, as the norm of $|\Psi\ran$ is fluctuating, but as long as $S_{Ob}$, $S_0$, and $S_2(\chi_{M'})$ are all large then we can reliably extract the expression 
\be
\label{eq:Renyi_topological_2}
\ol{e^{-S_2(\Psi_{M'})}}=e^{-S_2(\chi_{M'})}+e^{-S_2(\omega_{Ob'})}+e^{-2S_0}+\ldots,
\ee
which is accurate up to terms that involve products of the three quantities appearing here.

\subsection{JT gravity} \label{sec:4.3JT}

We will now consider closed universes in JT gravity in the presence of an observer. This theory has more structure than the topological model and reproduces the code calculation of section \ref{sec:code_generic}. Additionally, one new feature in JT is the appearance of a non-trivial QES in the closed universe which gives a geometric interpretation to the bottleneck found in the topological model. This bottleneck QES determines under what conditions operator reconstruction is possible in the closed universe.

The Euclidean action for JT gravity with asymptotically AdS boundaries is given by
\be
I_{JT}[g,\Phi] = -S_0 \chi(\mathcal{M}) - \frac{1}{2} \int_{\mathcal{M}} \sqrt{g} \Phi (R+2) - \int_{\partial M} \sqrt{h} \Phi (K-1)\,,
\ee
in terms of a dilaton field $\Phi$ and a two-dimensional metric $g$. We will be more precise about the boundary conditions in a moment. 

\begin{figure}
    \centering
    \includegraphics[width=0.4\linewidth]{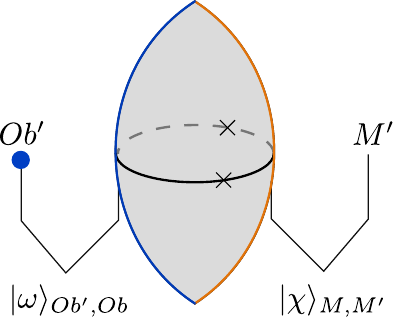}
    \caption{A big bang/big crunch closed universe in Lorentzian JT gravity with an observer (blue) and additional matter (orange). We clone the observer out to reference $Ob'$ and entangle the matter to an external reference $M'$. After taking into account the backreaction of the matter on the dilaton, two QESs are generated denoted by `\textbf{x}'. These QESs give a fundamental bottleneck on the observers Hilbert space in a closed universe.}
    \label{fig:jt_closed_universe}
\end{figure}

The Lorentzian theory for closed universes can be studied by ignoring the boundary term and taking the standard analytic continuation of the action $t_E \to i t_L$. There are classical solutions for Lorentzian big bang/big crunch closed universes on manifolds $\mathcal{M}= S^1 \times \mathbb{R}$. The solutions are given by $ds^2 = - dt^2 + b^2 \cos^2(t) d \sigma^2\,,\, \Phi = \Phi_c \sin(t)\,,$ where the spatial circle is periodic $\sigma \sim \sigma + 1$. These solutions describe a universe created from nothing at $t=-\pi/2$ which expands to a maximal size $b$ before re-collapsing at $t=\pi/2$, see figure \ref{fig:jt_closed_universe}. To include an observer in the closed universe we will add dynamical matter to the theory given by worldlines with mass $m_{Ob}$ and $m_M$ for observer and reference matter respectively.\footnote{The presence of these worldlines in the Lorentzian signature closed universe backreacts on the profile of the dilaton, but leaves the geometry rigid AdS$_2$. See \cite{Usatyuk:2024mzs} for some explicit constructions.} Technical details and all calculations are left to appendix \ref{app:RT}.

To introduce an observer into the closed universe we again endow the worldlines with extra flavor indices and entangle them with external reference systems $Ob'$ and $M'$ 
\be \label{eqn:JTstate_effective}
|\psi\ran_{\text{eff}} = \sum_{\substack{a,b \\i,j}} \omega_{a b} \chi_{i j}|a\ran_{Ob'} \otimes |\psi_{b,i}\ran \otimes |j\ran_{M'}\,,
\ee
where we have emphasized that in the code picture this state should be thought of as living in the effective description. The effective bulk state $|\psi_{a,i}\ran$ is a state of the closed universe in the canonically quantized theory with an observer in an internal state $a$ and additional matter in state $i$, see figure \ref{fig:jt_closed_universe} for a Lorentzian interpretation of this state.\footnote{In canonically quantized JT gravity the state of a closed universe can be specified by a wavefunctional $\psi(b) = \lan b | \psi\ran$ for the maximal geodesic length the universe attains. With the addition of matter fields the wavefunctions also must take as input a profile of appropriate matter fields on the Cauchy slice $\psi(b,\phi_1,\ldots)$.} 

In the path integral an insertion of the bulk effective state $|\psi_{i,a}\ran$ inserts a Euclidean boundary condition given by a geodesic circle of some specified length $b$ with observer and matter worldlines of specified flavors at determined locations along the circle. These boundary conditions are inconvenient for calculations, and as explained in \cite{Usatyuk:2024mzs} we can instead use boundary conditions given by asymptotic boundaries with matter operator insertions to generate closed universe states with wavefunctions highly peaked around a geodesic throat size $b$ determined by the details of the asymptotic conditions. When evaluating inner products using the path integral we thus make the replacement $|\psi_{i,a}\ran \to$ (Asymptotic circle of length $\beta+\beta'$ with Hermitian operators $\mathcal{O}_a, \Phi_i$ separated by $\beta,\beta'$) at the level of path integral boundary conditions.  

An example demonstrating these rules is to compute the average of the inner product of states with different matter and observer labels as we did in the topological model. We get
\be \label{eqn:JT_inner_prod}
\ol{\lan \psi_{i,a}| \psi_{j,b} \ran}=\delta_{ij} \delta_{a b} Z_1 = \raisebox{-10mm}{\includegraphics[scale=1.6]{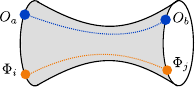}}\quad \,,
\ee
where the dotted lines denote geodesics connecting pairs of operators. In appendix \ref{app:RT} we explain how to evaluate these amplitudes in JT gravity, with $Z_1\sim(e^{S_0})^0$ given by \eqref{eqn:Z1} where we have given the scaling with the topological factor.

Instead of computing the variance in the inner product we will directly jump to computing the second Renyi entropy of the matter reference $M'$ after tracing out the rest of the system in \eqref{eqn:JTstate_effective}. The calculation is very similar to the topological model with the result\footnote{As explained in the case of the topological model, we must make sure the state is normalized to compute the Renyi entropy. In the limit that all the entropies and topological suppression are large we can get the leading order answer for the second Renyi by dividing by $(\ol{\Tr \Psi_{M'}})^2=Z_1^2$ which is the square of two Cylinders.}
\begin{align}
\label{eq:JT_entropy}
\ol{e^{-S_2(\Psi_{M'})}} =&~e^{-S_2(\chi_{M'})}+e^{-S_2(\omega_{Ob'})} + \Tr(\chi_M \chi_M^T) \Tr(\omega_{Ob} \omega^T_{Ob}) + \nonumber\\
+ & ~\frac{Z_2}{Z_1^2} + \mathcal{O}\left(e^{-2 S_0-S_2(\chi_{M'})}, e^{-2 S_0-S_2(\omega_{Ob'})},\ldots\right)
\end{align}

\begin{figure}
    \centering
    \includegraphics[width=0.4\linewidth]{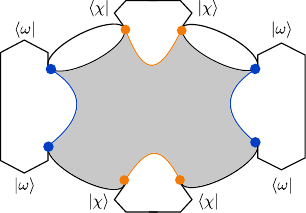}
    \caption{Dominant contribution to the fully connected second Renyi entropy. The observer and matter worldlines connect as indicated in the figure, giving a contribution that scales as $e^{-2 S_0} \Tr (\chi)^2 \Tr(\omega)^2 = e^{-2 S_0}$.}
    \label{fig:2ndrenyi_JT}
\end{figure}
The first line is from the three cylinder geometries. The second line comes from fully connected bulk geometries, where both the observer worldline and matter worldline are free to connect in all possible ways. In the limit that all the entropies and $S_0$ are large the most important connected contribution is given by observer and matter worldlines connecting as shown in figure \ref{fig:2ndrenyi_JT} which we denote by $Z_2 \sim e^{-2 S_0}$ and compute in \eqref{eqn:Znapp}. The most important terms are the first two and $Z_2/Z_1^2$, all others are subleading and exponentially suppressed in sums of $S_2(\chi_{M'}),S_2(\omega_{Ob'}), 2S_0$. We therefore have
\be
S_2(\Psi_{M'})\approx \text{min} \Big[ S_2(\chi_{M'}), S_2(\omega_{Ob'}),  \underbrace{\log\left(\frac{Z_1^2}{Z_2}\right)}_{2 S_0 + \ldots} \Big]\,,
\ee
where the last term dominates when both of the second Renyi entropies of the reference systems are much larger than $2 S_0$. The last term gives a fundamental bottleneck on the size of the observer's Hilbert space in a closed universe.  

We find a complete match between an observer in a closed universe in JT gravity with the expectations from the code. The first two terms match the calculation of the simple code in equation \eqref{codeobrenyi}, and the bottleneck can be seen in a code with more structure we study in appendix \ref{app:structuredcode}. 
\begin{figure}[H]
    \centering
    \includegraphics[width=1.1\linewidth]{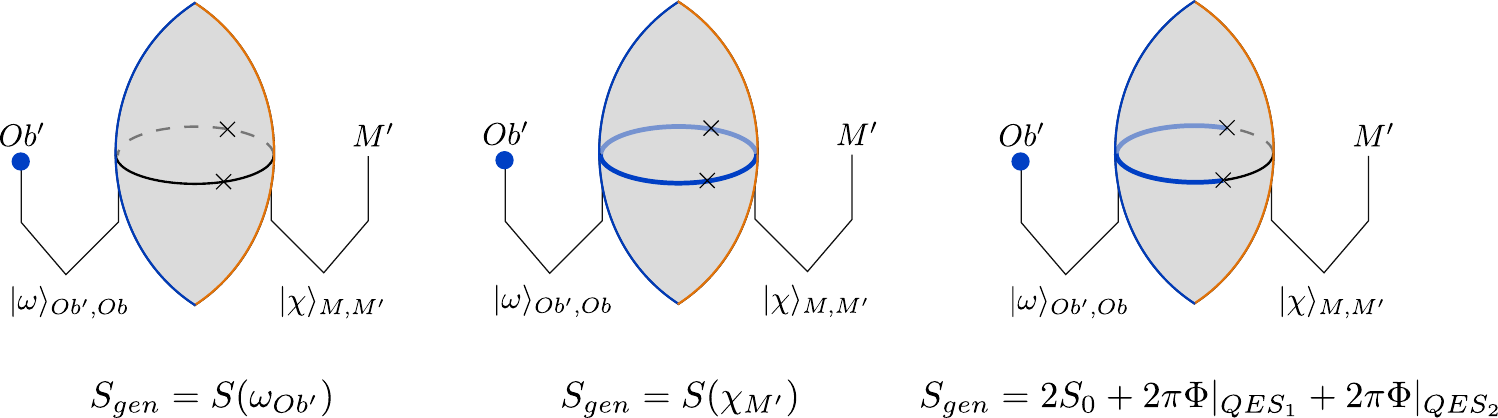}
    \caption{The three different quantum extremal surfaces for the entropy of the cloned observer system Ob'. The bulk entanglement wedge is denoted by the blue circle. In the first case the entanglement wedge contains $Ob'$. In the second case the entire closed universe, while in the third case the entanglement wedge ends at two QESs `\textbf{x}' and excludes the matter system $M'$.}
    \label{fig:JT_QES}
\end{figure}

It's also interesting to consider the von Neumann entropy of the cloned observer $Ob'$ and the under what conditions we can reconstruct operators acting inside the closed universe on $\mathcal{H}_{Ob'}$. The three phases for the QES are displayed in figure \ref{fig:JT_QES}. The entropy is given by
\be
S_{vN}(\omega_{Ob'}) = \text{min}\Big[ S(\omega_{Ob'}), S(\chi_{M'}), 2 S_0 + 2\pi \Phi\rvert_{QES_1} +2\pi\Phi\rvert_{QES_2} \Big]\,.
\ee
The last term is the bottleneck and comes from the two quantum extremal surfaces that surround the matter in figure \ref{fig:JT_QES}, these QESs are generated after backreaction of the matter is included. We use the replica trick to compute their value in appendix \ref{app:RT}. For the simplicity of this section we treat the bottleneck as approximately given by $2 S_0$.

The first case is the situation where  $S(\omega_{Ob'}) < S(\chi_{M'})$ and $S(\omega_{Ob'})< 2 S_0$, so the observer has less entropy than the matter that will be observed. In this case the map $\hat{V}$ is non-isometric and the information about the matter is encoded in $M$ instead of $Ob'$.  In fact the cloned observer $Ob'$ doesn't even have the observer $Ob$ in their entanglement wedge! In this situation we think the information accessible by $Ob'$ does not reflect the information observable by $Ob$.

The second case is when the observer entropy is $S(\chi_{M'}) < S(\omega_{Ob'})$ with $S(\chi_{M'}) < 2 S_0$. This is the situation of an observer looking at a small system in a big universe. The entanglement wedge of $Ob'$ is the entire closed universe, regardless of whether $S(\omega_{Ob'})< 2 S_0$ or $S(\omega_{Ob'})> 2 S_0$. In this case operators that act on the matter $M$ in the closed universe can be reconstructed on the cloned observers Hilbert space $Ob'$, and in fact the reconstruction is linear since the map $\hat{V}$ is close to being an isometry. 

When both the observer and matter exceed the bottleneck,  $S(\omega_{Ob'}) > 2 S_0$ and $S(\chi_{M'}) > 2 S_0$, the entanglement wedge of $Ob'$ contains only a subset of the closed universe and in particular no longer contains the matter system. $Ob'$ loses the ability to reconstruct operators acting on the matter. Interestingly, this failure occurs even when the observer's entropy is much larger than the matter entropy $S(\omega_{Ob'}) \gg S(\chi_{M'})$ and so the matter is a `small' subsystem relative to the observer.

\subsection{A microscopic model}\label{ETHsec}

The gravitational theories we have studied so far at best have dual holographic interpretations as ensemble averages of theories \cite{Saad:2019lba,Marolf:2020xie,Jafferis:2022wez}.  It is not clear if gravity in higher dimensions has such an interpretation, and in any event it would be nice to have a version of our story which works in a fixed microscopic theory that has a sensible Einstein gravity limit.   In this section we explain how to use the Eigenstate Thermalization Hypothesis (ETH) to reproduce our results from the code and path integral approaches within a fixed closed universe theory that is dual to some particular holographic CFT such as $\mathcal{N}=4$ super Yang-Mills theory.  We will see that this allows us to have a concrete formula for the encoding map $\hat{V}$ in terms of CFT quantities, and to obtain the same results as before for the inner product and Renyi entropy.  For previous work using ETH in a similar manner see \cite{Pollack:2020gfa,Belin:2020hea}.

The Eigenstate Thermalization Hypothesis (ETH) roughly speaking says that in generic many-body quantum systems simple observables show thermal behavior for generic initial states \cite{Srednicki:1994mfb}. More precisely, it says that the matrix elements of a simple hermitian operator $A$ in energy eigenstates obeys
\begin{align}
\label{eq:ETH}
    \bra{E_{\alpha}}A\ket{E_{\beta}} = \mathcal{A}(\bar E_{\alpha\beta})\delta_{\alpha\beta}+e^{-S(\bar E_{\alpha\beta})/2}f^A(E_{\alpha},E_{\beta})R^A_{\alpha\beta},
\end{align}
with $\bar E_{\alpha\beta}\equiv\frac{E_{\alpha}+E_{\beta}}{2}$. Here $\mathcal{A}$ and $f^A$ are smooth functions of $E$, while $R^A_{\alpha,\beta}$ depends sensitively on $\alpha$ and $\beta$ with rapidly varying phase.  
When we sum over a large number of energy eigenstates, we can treat $R$'s as Gaussian random variables satisfying
\begin{align}
\label{eq:ETH_Gaussian}
    \overline{R^A_{\alpha\beta}}=0,\ \ \ \overline{R^A_{\alpha\beta}R^A_{\gamma\delta}} = \delta_{\alpha\delta}\delta_{\beta\gamma}
\end{align}

To construct closed universe states with observers, we define $V$ in figure \ref{Ocode2fig} by taking a thermal trace: 
\begin{align}
	\hat V\ket{i}_M =\ \sum_{a,b}\omega_{ab}\ket{a}_{Ob'}\Tr(e^{-\beta H/2}\mO_be^{-\beta H/2}\Phi_i)
\end{align}
where $\mathcal{O}_b$ inserts an observer in classical state $b$ and $\Phi_i$ inserts matter particle in state $i$. 
The inner product is given by
\begin{align}
	\bra{j}\hat V^{\dagger}\hat V\ket{i} =\sum_{b,c}\Tr(e^{-\beta H/2} \Phi_j^{\dagger}e^{-\beta H/2}\mO_c^{\dagger}) (\omega_{Ob})_{cb}\Tr(e^{-\beta H/2}\mO_be^{-\beta H/2}\Phi_i)
\end{align}

To simplify, let's assume both $\mO$ and $\Phi$ are Hermitian operators with vanishing thermal one point functions. Using ETH hypothesis in \eqref{eq:ETH}, we have
\begin{align}
	\Tr(e^{-\beta H/2}\mO_be^{-\beta H/2}\Phi_i) =\ & \sum_{\alpha,\beta}e^{-\beta\bar E_{\alpha\beta}}e^{-S(\bar E_{\alpha\beta})}f^{\mO_b}(E_{\alpha}, E_{\beta})f^{\Phi_i}(E_{\beta}, E_{\alpha})R^{\mathcal{O}_b}_{\alpha\beta}R^{\Phi_i}_{\beta\alpha}
\end{align}
To match the rather simple matter theory we have discussed above we will further assume the that function $f^{\Phi_i}$ is independent of $i$.  Similarly we will assume that $f^{\mO_b}$ is independent of $b$. These assumptions will not be literally true for more realistic matter, but we don't expect this dependence to lead to anything interesting in this context. To simplify notation, we also let $f_{\alpha\beta} \equiv f(E_{\alpha},E_{\beta})$. Now we can use \eqref{eq:ETH_Gaussian} to deal with the rapidly varying phases, which gives
\begin{align}
\label{eq:ETH_average_0}
	\overline{\bra{j}\hat V^{\dagger}\hat V\ket{i}}  = \delta^{ij}(\Tr\omega_{Ob})\qty(\sum_{\alpha,\beta}e^{-2\beta \bar E_{\alpha\beta}}e^{-2S(\bar E_{\alpha\beta})}\qty|f^{\mO}_{\alpha\beta}|^2\qty|f^{\Phi}_{\alpha\beta}|^2)
\end{align}
We note that what's inside the parentheses in \eqref{eq:ETH_average_0} is of order one, which corresponds to the fact that cylinder diagram in gravitational path integral gives contribution of order one. We write a bar here to resemble our path integral notation and remind us that we used the ETH approximation, but we emphasize that we have only used \eqref{eq:ETH_Gaussian} to approximately evaluate two of the sums: we are \textit{not} averaging over theories.  

Now we consider the square of the inner product. Using \eqref{eq:ETH_Gaussian} and including all nine different ways of contracting the $R$,\footnote{There is an exact correspondence between these nine different ways of doing ETH contractions and the nine different ways to contract orthogonal matrices in the structured code calculation. See appendix \ref{app:structuredcode} for more details.} we obtain
\begin{align}
\label{eq:ETH_average_1}
	&\overline{|\bra{i}\hat V^{\dagger}\hat V\ket{j}\bra{j}\hat V^{\dagger}\hat V\ket{i}|}\nonumber\\	
	=\ &\qty(\sum_{\alpha,\beta}e^{-2\beta\bar E_{\alpha\beta}}e^{-2S\bar E_{\alpha\beta}}\qty|f^{\mO}_{\alpha\beta}|^2\qty|f^{\Phi}_{\alpha\beta}|^2)\qty[\delta_{ij}\qty(1+\Tr(\omega_{Ob}\omega_{Ob}^T))+\Tr(\omega_{Ob}^2)]\nonumber\\
	&+\qty(\sum_{\alpha,\beta}e^{-4\beta\bar E_{\alpha\beta}}e^{-4S(\bar E _{\alpha\beta})}\qty|f^{\mO}_{\alpha\beta}|^4\qty|f^{\Phi}_{\alpha\beta}|^4)\qty[\qty(1+\delta_{ij})\qty(1+\Tr(\omega_{Ob}\omega_{Ob}^T))+2\delta_{ij}\Tr(\omega_{Ob}^2)]
\end{align}
 The second line on the right hand side of \eqref{eq:ETH_average_1} is suppressed by $e^{-2S(E)}$ which corresponds to the topological suppression in gravitational path integral calculations. 
We emphasize that although we are not averaging, the ETH approximation picks up terms in the square of the inner product which it neglects for the inner product. This is because it only includes terms where the phases in $R$ cancel, and there are more such terms when we take the absolute value squared.  
 Indeed defining
\begin{align}
&Z_1 = \sum_{\alpha,\beta}e^{-2\beta\bar E_{\alpha\beta}}e^{-2S(\bar E_{\alpha\beta})}\qty|f^{\mO}_{\alpha\beta}|^2\qty|f^{\Phi}_{\alpha\beta}|^2\label{eq:Z1}\\
&Z_2 = \sum_{\alpha,\beta}\qty(e^{-2\beta\bar E_{\alpha\beta}}e^{-2S(\bar E _{\alpha\beta})}\qty|f^{\mO}_{\alpha\beta}|^2\qty|f^{\Phi}_{\alpha\beta}|^2)^2\label{eq:Z2}
\end{align}
and assuming $\omega$ is $\CRT$ invariant, we have
\begin{align}
\label{eq:fluctuation_ETH}
&\overline{|\bra{j}\hat V^{\dagger}\hat V\ket{i}-\overline{\bra{j}\hat V^{\dagger}\hat V\ket{i}}|^2}
\nonumber\\
=\ &Z_1^2
\left[\qty(1+\delta_{ij})\qty(e^{-S_2(\omega_{Ob})}+\frac{Z_2}{Z_1^2})+(1+3\delta_{ij})e^{-S_2(\omega_{Ob})}\frac{Z_2}{Z_1^2}\right]
\end{align}
The overall factor $Z_1^2$ has to do with the fact that states $\hat{V}\ket{i}$ are not quite normalized. Note that the inner product fluctuation is suppressed by the observer entropy $e^{-S_2(\omega_{Ob})}$ as long as $S_2(\omega_{Ob})<\log(\frac{Z_1^2}{Z_2})\sim 2S(E)$. 

We can again entangle the matter to a reference system $M'$ in a state $\chi$ and compute the second Renyi entropy of the matter reference system. Using the same method we find
\begin{align}
\label{eq:Renyi_ETH}
    \frac{1}{Z_1^2}\overline{\Tr(\Psi_{M'}^2)} =& \Tr(\chi_M^2)+\Tr(\omega_{Ob}^2)+\Tr(\omega_{Ob}\omega_{Ob}^T)\Tr(\chi_M\chi_M^T)\nonumber\\
	&\ +\frac{Z_2}{Z_1^2}\Big[1+\Tr(\chi_M\chi_M^T)+\Tr(\omega_{Ob}\omega_{Ob}^T)\nonumber\\
    &\ \ \ +\Tr(\omega_{Ob}\omega_{Ob}^T)\Tr(\chi_M^2)+\Tr(\omega_{Ob}^2)\Tr(\chi_M^2)+\Tr(\omega_{Ob}^2)\Tr(\chi_M\chi_M^T)\Big]
\end{align}
The second Renyi entropy satisfies
\begin{align}
\label{eq:entropy_ETH}
\overline{e^{-S_2(\Psi_{M'})}} = e^{-S_2(\chi_{M'})}+e^{-S_2(\omega_{Ob'})}+\frac{Z_2}{Z_1^2}+\text{subleading terms}
\end{align}
where $\frac{Z_2}{Z_1^2}$ is of order $e^{-2S(E)}$.

Equation \eqref{eq:entropy_ETH} matches our results \eqref{eq:Renyi_topological_2} in topological model and \eqref{eq:JT_entropy} in JT gravity. There is also a detailed match between \eqref{eq:fluctuation_ETH}, \eqref{eq:Renyi_ETH} and our expressions \eqref{eq:code_structured}, \eqref{eq:code_structured_Renyi} from the structured code.

We would like to emphasize that the calculations in this section were carried out in a fixed microscopic theory using the ETH approximation. This is a completely different perspective from the gravitational path integral in JT gravity, which is dual to an ensemble average of theories. Nevertheless we saw detailed correspondence between ETH expressions \eqref{eq:Z1},\eqref{eq:Z2},\eqref{eq:entropy_ETH} and the JT expressions \eqref{eqn:Z1}, \eqref{eqn:Znapp}, \eqref{eq:JT_entropy}. It has been argued that there is no ensemble averaging of theories in higher dimensions \cite{McNamara:2020uza}, so the ETH may give a more robust interpretation of the path integral in higher dimensions.

\section{An observer in a black hole}\label{bhobsec}
In this section we explain how explicitly including an observer can sometimes be important for understanding the emergence of the black hole interior.  One way to get into a regime where this is true is for an observer to fall into a large black hole which then evaporates for a long time, after which we perform some complete measurement on the Hawking radiation.  This puts the black hole in a pure quantum state, and the Hilbert space of such states can be quite small if the remaining black hole is small.  How then are we supposed to describe the experiences of the observer who fell in?  Scenarios which are essentially the same as this one have been discussed before in \cite{Kourkoulou:2017zaj,Almheiri:2018xdw,Antonini:2024mci}.  The essential point is that if we now allow the black hole to completely evaporate, this observer is essentially just living in a closed universe.  We we have already seen that it is necessary to include the observer explicitly for a closed universe, so this scenario gives us a way to interpolate between our closed universe scenario where the observer must be included and the standard one where an observer jumps into a large black hole and no explicit observer is necessary.  In this section we explore the details of this interpolation from both the coding and the path integral point of view.  

\subsection{Review of non-isometric codes for the black hole interior}
\bfig
\includegraphics[height=5cm]{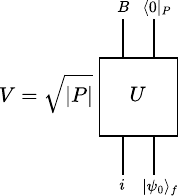}
\caption{A simple code for the emergence of the black hole interior.}\label{Ucodefig}
\efig
As mentioned in the introduction, we can describe the emergent spacetime inside of a black hole using a linear but not necessarily isometric encoding map 
\be
V:\mathcal{H}_i\to \mathcal{H}_B
\ee
from the Hilbert space $\mathcal{H}_i$ of effective field theory modes in the black hole interior to a Hilbert space $\mathcal{H}_B$ of black hole microstates.  A simple model of this map was given in \cite{Akers:2022qdl}: we take the interior state $|\psi\ran_i$, tensor it with some fixed state $|\psi_0\ran_f$, act on the combined system with a generic unitary $U$, and then project a subset $P$ of the final degrees of freedom onto a fixed state $\lan 0|_P$.  See figure \ref{Ucodefig} for an illustration. The reason that $U$ is not required to be orthogonal is that this is a model for a black hole in a spacetime with an asymptotic boundary, so the physical states do not need to be $\CRT$-invariant.  This map approximately preserves the inner product of interior states up to errors which are of order $\frac{1}{\sqrt{|B|}}$.  More concretely we have\footnote{Unitary integrals are a bit easier to compute than orthogonal integrals but basically work in the same way, see e.g. appendix A of \cite{Akers:2022qdl} for a refresher course.} 
\be
\int dU\Big|\lan \phi|V^\dagger V|\psi\ran-\lan \phi|\psi\ran\Big|^2=\frac{|B|^2|P|^2}{|B|^2|P|^2-1}\left(1-\frac{1}{|P|}\right)\left(1-\frac{|\lan \phi|\psi\ran|^2}{|B||P|}\right)\frac{1}{|B|},
\ee
which is in general is upper bounded by $2/|B|$ and in the convenient limit $|P|\to \infty$ simply reduces to 
\be
\int dU\Big|\lan \phi|V^\dagger V|\psi\ran-\lan \phi|\psi\ran\Big|^2=\frac{1}{|B|}.
\ee
This model captures many features of quantum black hole physics, and in particular if we split the interior system $i$ into left and right moving modes $\ell$ and $r$, and entangle $r$ with an external system $R$ representing the Hawking radiation and $\ell$ with an external reference system $L$, then the encoded state
\be
|\Psi\ran_{BR}=(V\otimes I_R)\left(|\psi\ran_{\ell,L}\otimes|\chi\ran_{r,R}\right)
\ee
has a radiation entropy 
\be
S(\Psi_R)=\min\left[S(\chi_R),\log |B|+S(\psi_\ell)\right]
\ee
which agrees with the quantum extremal surface result from \cite{Penington:2019npb,Almheiri:2019psf} provided that we interpret $\log |B|$ as the horizon area divided by $4G$.  In the code this expression also holds for the $n$th Renyi entropy, which reflects the fact that the code is most closely analogous to ``fixed-area'' states in gravity \cite{Akers:2018fow,Dong:2018seb}.    

One feature of this code however is that when the black hole is mostly evaporated, $|B|$ is small so the inner product is not well-preserved by $V$.  In \cite{Akers:2022qdl} this was interpreted as a positive feature, since an observer who jumps into a small black hole should not expect a valid semiclassical experience.  What should we say however about the scenario described above, where the observer jumps in and then the black hole evaporates down to small size?  We will now see that by including the observer explicitly in the code we can get a reasonable description of their experiences up to errors which are exponentially small in $S_{Ob}$ even when the black hole is small. 

\subsection{Including the observer in the black hole code}
As in our discussion of a closed universe, we introduce a modified code $\hat{V}:\mathcal{M}\to \mathcal{H}_B\otimes \mathcal{H}_{Ob'}$ which maps us from the effective field theory degrees of freedom in the black hole interior not including the observer to the combination of the black hole microstates $B$ and the cloned observer $Ob'$.  See figure \ref{BHcodefig} for an illustration.  Computing the average fluctuation of the inner product we get
\be
\int dU |\lan \phi|\hat{V}^\dagger \hat{V}|\psi\ran-\lan \phi|\psi\ran|^2=\frac{|B|^2|P|^2}{|B|^2|P|^2-1}\left(1-\frac{1}{|P|}\right)\left(e^{-S_2(\omega_{Ob'})}-\frac{|\lan \phi|\psi\ran|^2}{|B||P|}\right)\frac{1}{|B|},
\ee
which is upper bounded by $2\frac{e^{-S_2(\omega_{Ob'})}}{|B|}$ and in the limit of $|P|\to \infty$ just becomes
\be
\label{eq:BH_fluctuations_code}
\int dU |\lan \phi|\hat{V}^\dagger \hat{V}|\psi\ran-\lan \phi|\psi\ran|^2=\frac{e^{-S_2(\omega_{Ob'})}}{|B|}.
\ee
Thus we see that fluctuations of the inner product are now suppressed by $e^{-S_{Ob}}$ as well as $\frac{1}{|B|}$, so a big observer can have a valid semiclassical experience in a black hole even if we wait until the black hole is almost/completely evaporated to decode it.

\bfig
\includegraphics[height=5cm]{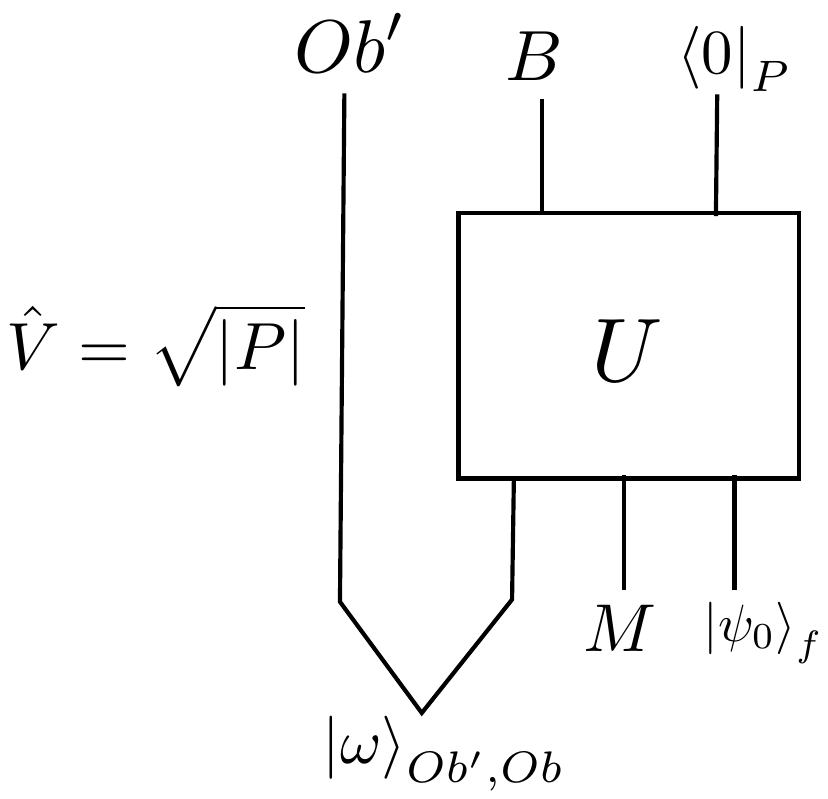}
\caption{A simple code for an evaporating black hole with an observer.}\label{BHcodefig}
\efig

We can also consider the second Renyi entropy of a matter reference system $M'$ which is entangled with $M$ in a state $|\chi\ran_{MM'}$.  Mathematically this is identical to the Page curve calculation described in the previous subsection, we just need to make the replacements $R\to M'$, $r\to M$, $\ell \to Ob$, $L\to Ob'$, so the second Renyi entropy is
\be
\label{eq:BH_Renyi_code}
S_2(\Psi_{M'})=\min \left[S_2(\chi_{M'}),\log |B|+S_2(\omega_{Ob'})\right].
\ee
Thus we see that even when the black hole is small there can still be sizable entanglement between its interior matter and a reference system, as long as there is a big observer who is also inside.  

\subsection{Including the observer in JT gravity}
\begin{figure}
  \centering
  \includegraphics[width = 3.6in]{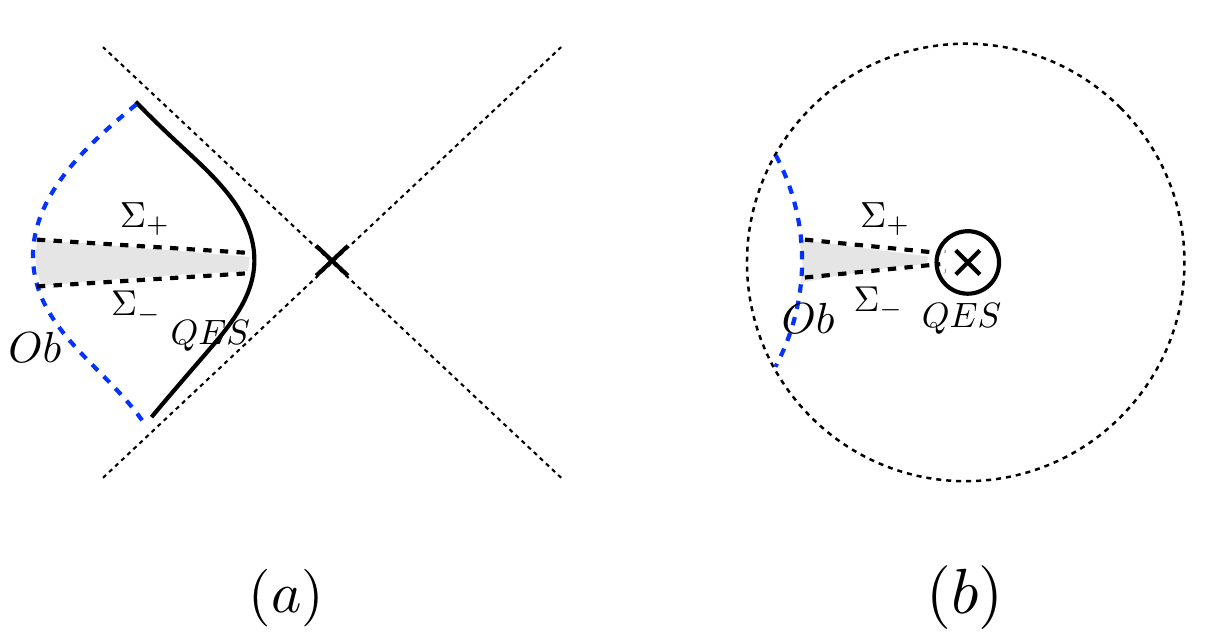}
  \caption[Caption for LOF]{Different types of boundaries for JT gravity in the island. (a) is in Lorentzian signature while (b) is in Euclidean signature. The crosses are true quantum extremal surfaces. Regulated boundaries $\Gamma_{QES}$ sitting close to the true quantum extremal surfaces are represented by solid black lines. The end of world brane boundaries modeling the observer $\Gamma_{Ob}$ are denoted by dashed blue lines. The remaining boundaries $\Sigma_{\pm}$ giving the Cauchy surfaces connecting the QES and observer are represented by dashed black lines. }
  \label{fig:JT_boundary}
\end{figure}

We now explain how to reproduce the code results of the previous subsection using the path integral in JT gravity.   In the classic treatment of an evaporating black hole in this theory the interior region is part of a larger spacetime that includes the black hole exterior and bath into which the black hole is radiating \cite{Almheiri:2019psf}.  To more closely parallel our discussion of the closed universe we want to focus on the interior only, for example in a pure state obtained by doing a complete projective measurement on the radiation, so we introduce a new kind of boundary condition for JT that allows us to view the interior as a closed dynamical system in its own regard.  The idea is to consider a spatial interval which is bounded on one side by a quantum extremal surface of fixed dilaton and on the other side by an end-of-the world brane representing the observer.  We will see that the dilaton at the extremal surface plays the role of $\log |B|$ in the code.  We will take the Lorentzian action between a future Cauchy surface $\Sigma_{+}$ and a past Cauchy surface $\Sigma_-$ to be
\begin{align}\nonumber
I=&\frac{\Phi_0}{2}\left[\int_\mathcal{M} d^2 x \sqrt{-g}R+2\int_{\Gamma_{QES}+\Gamma_{Ob}}dx\sqrt{-h}K\right]+\frac{1}{2}\int_\mathcal{M} d^2x \sqrt{-g}\Phi\left(R+2\right)\\
&+\int_{\Gamma_{QES}}dx\sqrt{-h}\Phi(K+\alpha)+\int_{\Gamma_{Ob}}dx\sqrt{-h}\left(\Phi K-\mu\right)+\int_{\Sigma_+-\Sigma_-}dx\sqrt{h}\Phi K,\label{JTL}
\end{align}
with boundary conditions
\begin{align}
K|_{\Gamma_{Ob}}=0,\ \ \ \nabla_n\Phi|_{\Gamma_{Ob}} = \mu\,,
\end{align}
on the observer wordline and
\begin{align}
\Phi|_{\Gamma_{QES}} = \Phi_e,\ \ \ \nabla_{n}\Phi|_{\Gamma_{QES}} = -\alpha \Phi,\ \ \alpha>0\,,
\end{align}
at the regulated QES boundary. With these boundary conditions the action \eqref{JTL} is stationary at solutions of the JT equations of motion up to terms on $\Sigma_{\pm}$, and with our choice of boundary terms on $\Sigma_{\pm}$ if we require these variations to also vanish this sets
\begin{align}
\label{eq:JT_QES_bc}
	K\delta\Phi|_{\Sigma_{\pm}} = 0,\ \ \ \qty(\nabla_n\Phi)h^{\mu\nu}\delta g_{\mu\nu}|_{\Sigma_\pm}=0.
\end{align}
To get the regulated QES boundary to coincide with the actual QES, as we want to describe the island, we must take the limit $\alpha\to 0^+$. See figure \ref{fig:JT_boundary} for an illustration.    

To compute entropies and fluctuations we want to study this theory in Euclidean signature. The Euclidean action is given by\footnote{$S_0 = 2\pi\Phi_0$.} 
\begin{align}\nonumber
I=&-2\pi \Phi_0 \chi(\mathcal{M})-\frac{1}{2}\int_\mathcal{M} d^2x \sqrt{g}\Phi\left(R+2\right)\\
&-\int_{\Gamma_{QES}}dx\sqrt{h}\Phi(K+\alpha)-\int_{\Gamma_{Ob}}dx\sqrt{h}\left(\Phi K-\mu\right)-\int_{\Sigma_+-\Sigma_-}dx\sqrt{h}\Phi K,\label{eq:JT_action}
\end{align}

\begin{figure}
  \centering
  \includegraphics[width = 4.6in]{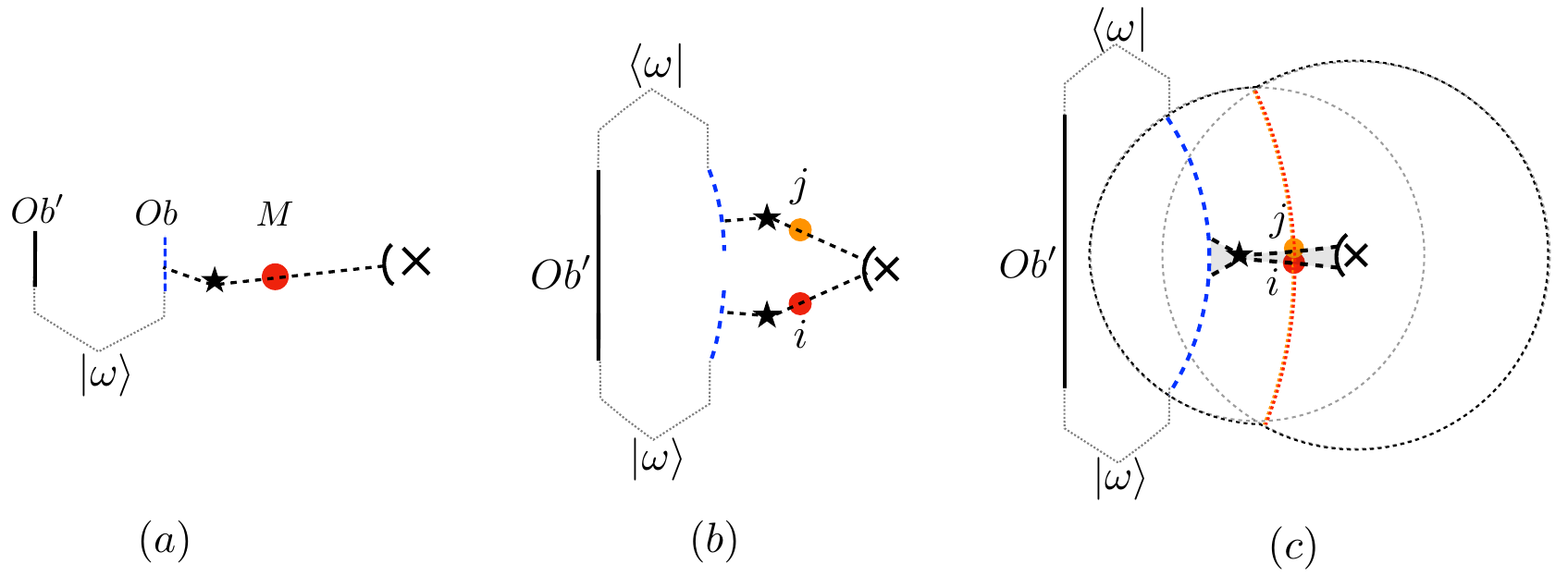}
  \caption{Computing the inner product between states with different matter species in presence of a quantum extremal surface and an observer. 
}
  \label{fig:JT_QES_inner_product}
\end{figure}

\begin{figure}
    \centering
\includegraphics[width=0.35\linewidth]{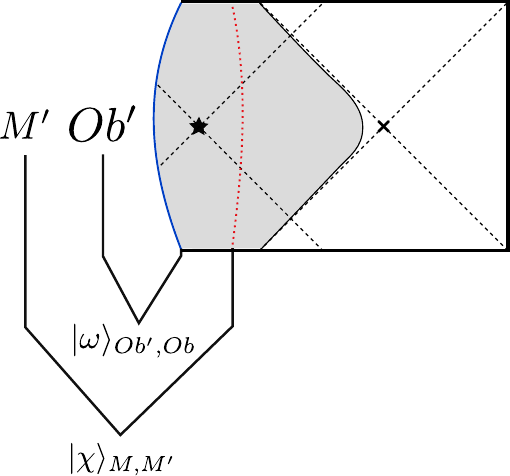}
    \caption{An example where the model of island in this section is embedded into a more complete Penrose diagram. The right portion of the diagram is included in the full spacetime, but we study the shaded region.}
    \label{fig:JT_Interior}
\end{figure}

To compute entropies we must also introduce worldline matter $M$ with different species $i$ which adds to the action \eqref{eq:JT_action} a term $I_M=\sum_i m\int ds\sqrt{g_{\mu\nu}\dot X_i^{\mu}(s)\dot X_i^{\nu}(s)}$. See figure (Figure \ref{fig:JT_QES_inner_product}(a)).\footnote{$X^{\mu}(s)$ is the worldline of the massive particle with s an affine time along the worldline.} With matter present, a new quantum extremal surface may be generated due to the backreaction of the matter, see for example \cite{Goel:2018ubv}. Instead of specifying the location of the matter and working out the backreaction we will specify the dilaton $\Phi_{e2}$ at the new quantum extremal surface. In figure \ref{fig:JT_QES_inner_product}, the new quantum extremal surface is labelled by a star. Figure \ref{fig:JT_Interior} shows an example where this model of the island with an observer and matter is embedded into a more complete Penrose diagram.

We begin by studying the inner product between states with different matter species. The boundary conditions when computing $\ol{\lan j | i\ran}$ require the regulated QES and $Ob'$ from the ket and bra to contract  (Figure \ref{fig:JT_QES_inner_product}(b)). The saddlepoint geometry is given by a disk topology that degenerates into an infinitesimal strip (Figure \ref{fig:JT_QES_inner_product}(c)).\footnote{For saddle configuration, the two geodesics slices will sit on top of each other and be both perpendicular the end of world brane. Easy to see this configuration satisfies the boundary condition given in \eqref{eq:JT_QES_bc}. The geometry will be degenerate and the contribution to the action comes entirely from corner terms. }
With this configuration we get
\begin{align}
    \overline{\bra{j}\ket{i}} = \delta_{ij}\exp\qty[\pi\qty(\Phi_0+\Phi_{Ob})+\pi\qty(\Phi_0+\Phi_e)]\,,
\end{align}
where $\Phi_{Ob}$ is the minimal value of dilaton on the end of world brane given by
\begin{align}
   \Phi_{Ob} = \sqrt{\mu^2+\Phi_{e2}^2}
\end{align}

\begin{figure}
  \centering
  \includegraphics[width = 5.6in]{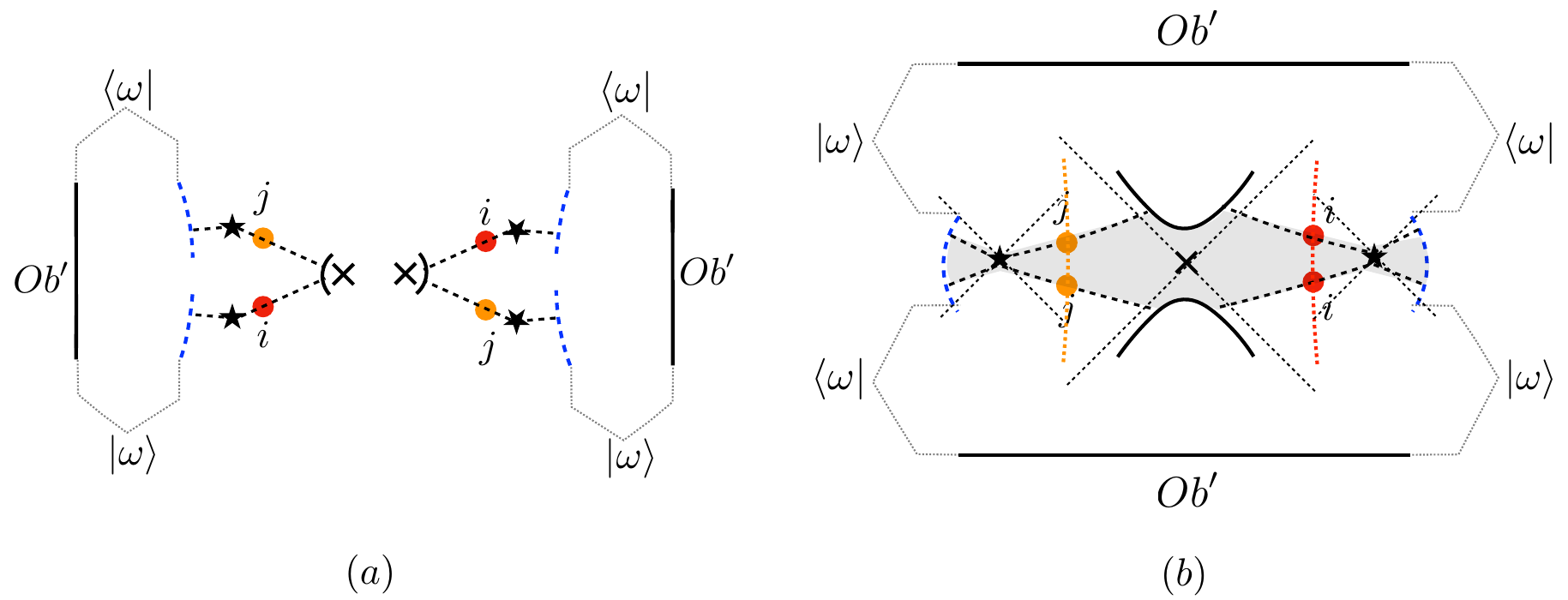}\\
  \includegraphics[width = 3.6in]{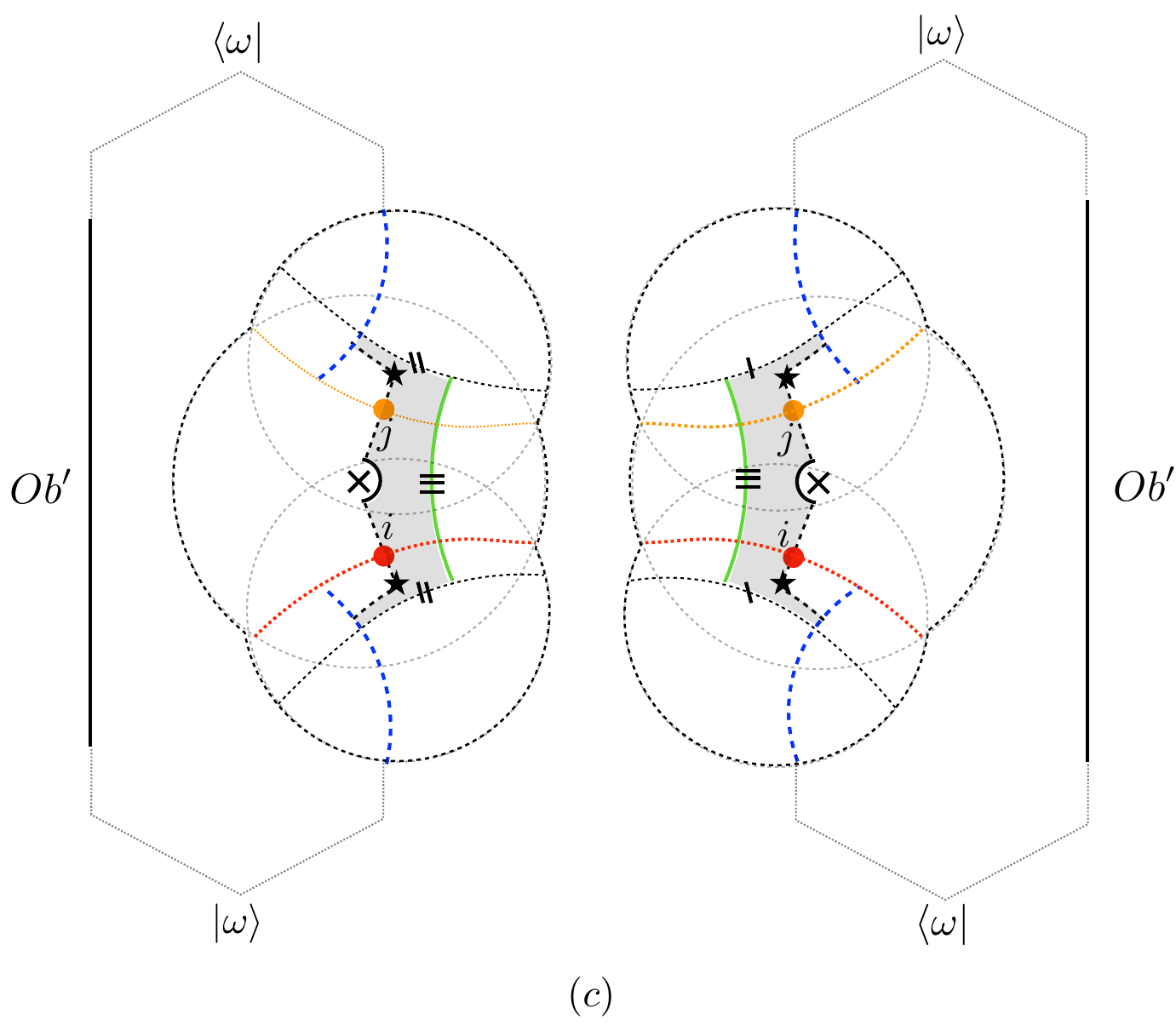}
  \caption{Path integral computation of the square of the inner product $\ol{|\lan j | i\ran|^2}$.}
  \label{fig:JT_QES_inner_product_square}
\end{figure}

Next we consider the path integral calculation of the square of the inner product (Figure \ref{fig:JT_QES_inner_product_square}(a)). At leading order, other than the contribution from two copies of Figure \ref{fig:JT_QES_inner_product}(c), we have two additional contributions (Figure \ref{fig:JT_QES_inner_product_square}(b) and (c)).
Figure \ref{fig:JT_QES_inner_product_square}(b) is a portion of the Lorentzian signature black hole. In the limit of $\alpha\rightarrow 0$, the dominant configuration is again a degenerate geometry. It gives a contribution of $\Tr(\omega_{Ob'}^2)\exp\qty[2\pi(\Phi_0+\Phi_{Ob})]$.
If we combine the results from figure \ref{fig:JT_QES_inner_product}(c) and figure \ref{fig:JT_QES_inner_product_square}(b), we obtain
\begin{align}
\label{eq:JT_inner_product_square_1}
    &\overline{|\bra{j}\ket{i}-\overline{\bra{j}\ket{i}}|^2}=e^{-S_2(\omega_{Ob'})}\exp\qty[-2\pi\qty(\Phi_0+ \Phi_e)]+\mathcal{O}(e^{-2S_0})\,.
\end{align}
Up to terms of order $e^{-2S_0}$, \eqref{eq:JT_inner_product_square_1} matches the code calculation in \eqref{eq:BH_fluctuations_code}. 

Figure \ref{fig:JT_QES_inner_product_square}(c) is a cylinder diagram which gives contribution $\exp(2\pi\Phi_{Ob}-2\pi\Phi_{e2})$ as $\alpha\rightarrow 0$. Its computation is in appendix \ref{app:JT_QES}. If we include contribution from this configuration, we will obtain
\begin{align}
\label{eq:JT_QES_fluctuation}
    &\overline{|\bra{j}\ket{i}-\overline{\bra{j}\ket{i}}|^2}=\exp\qty[-2\pi\qty(\Phi_0+\Phi_e)]\Big(e^{-S_2(\omega_{Ob'})}+\exp\qty[-2\pi\qty(\Phi_0+\Phi_{e2})]\Big)+\text{subleading}.
\end{align}

We can also entangle the matter with a reference system $M'$ in state $\chi$ and study the second Renyi entropy of $M'$. The result is
\begin{align}
    \overline{\Tr(\Psi_{M'}^2)}
    =&\ e^{-S_2(\chi_{M'})}\exp(4\pi\Phi_0+2\pi\Phi_e+2\pi\Phi_{Ob})+e^{-S_2(\omega_{Ob'})}\exp(2\pi(\Phi_0+\Phi_{Ob}))\nonumber\\
    &\ +\exp(2\pi(\Phi_{Ob}-2\pi\Phi_{e2}))+\text{subleading},
\end{align}
so the second Renyi entropy is approximately given by
\begin{align}
\label{eq:JT_QES_Renyi}
 S_2(\Psi_M') \approx \min\Big[S_2(\chi_{M'}),2\pi(\Phi_0+\Phi_e)+S_2(\omega_{Ob'}),2\pi(\Phi_0+\Phi_e)+2\pi(\Phi_0+\Phi_{e2})\Big].
\end{align}
We emphasize that equations \eqref{eq:JT_inner_product_square_1}-\eqref{eq:JT_QES_Renyi} are in the $\alpha\rightarrow 0$ limit. 
The third option in the minimum in equation \eqref{eq:JT_QES_Renyi} can be interpreted as the sum of areas of two quantum extremal surfaces. One is the QES where we set our boundary condition with area given by $\Phi_0+\Phi_e$. The other one is the new QES with area $\Phi_0+\Phi_{e2}$. One may worry that when both $\Phi_0+\Phi_e$ and $\Phi_0+\Phi_{e2}$ are small, the inner product fluctuation in \eqref{eq:JT_QES_fluctuation} will be large and the Renyi entropy in \eqref{eq:JT_QES_Renyi} will be small despite a large observer entropy. However, in this case the matter sits between two small QES's which can potentially compete with each other. In this case it is expected that the encoding of the matter into the observer $Ob'$ will be bad \cite{Hayden:2018khn}. In this paper we will avoid this regime. In fact, this subtly will not arise in cases involving actual black holes since if matter is thrown in too late after the observer jumps in the observer will not be able to see the matter.

On the other hand, if the question we want to ask is how much matter can the system hold with a fixed observer size $S(\omega_{Ob'})$ and a fixed QES with area $\Phi_0+\Phi_e$, we can always arrange the matter such that $\Phi_0+\Phi_{e2}$ is large, and the ultimate bottleneck will be given by $S(\omega_{Ob'})+2\pi(\Phi_0+\Phi_e)$ as predicted by the code calculation in \eqref{eq:BH_Renyi_code}. We see that when the quantum extremal surface is large, we can simply describe bulk physics though the holographic system. When it becomes too small, as in the case of a black hole that was almost completely evaporated and its radiation completely measured, introducing an observer becomes necessary.

\section{Discussion}\label{discussionsec}
In this final section we will make a few comments about the interpretation and generality of our theory of the observer in a closed universe.  
\bi
\item \textbf{Suppressing cross connections vs. forbidding them:} One way to think about the puzzling features of quantum gravity in a closed universe is that they arise from ``cross-connection'' geometries such as the second and third saddles in figure \ref{topip2fig}.  Mathematically what our introduction of $Ob'$ does is suppress these geometries by a factor of $e^{-S_{Ob}}$.  It is therefore tempting to take the limit $S_{Ob}\to \infty$, so that cross connections are forbidden altogether.  After all if you yourself are an observer in a closed universe, and you want to describe your own experiences, it is natural to take $S_{Ob}\to \infty$: in practice you never measure anything to precision $e^{-S_{Ob}}$, and even if you could you wouldn't be able to store the result in your brain.  There are two reasons why this limit requires some care.  The first is that if you and I are different observers in a closed universe, with $S_{me}\gg S_{you}$, then it is fair game for me to try to make sense of you with precision $e^{-S_{you}}$ and in that case I should include effects of this size.  Said differently, if I treat you as part of the matter system $M$ what I find should be consistent with what you find when you treat yourself as an observer up to effects which are exponentially small in $S_{you}$.  The second is that it is probably unphysical to really take $S_{Ob}$ to infinity without also taking the observer mass to infinity, and this can have catastrophic effects on the cosmology due to gravitational backreaction unless we also take $G\to 0$ at the same time.  Taking this limit consistently therefore suppresses quantum gravity effects, while we would like to be able to consider an observer who is robust enough to learn nontrivial things about quantum gravity.  In the context of the topological model or JT gravity, this is the regime $S_{Ob}\gg S_0\gg 1$.  In this limit the backreaction of the observer is important but not necessarily catastrophic (depending on what else is going on in the cosmology), and we'd like to have a theory that describes it.   On the other hand we'd also like to see that when $S_{Ob}\ll S_0$ the precision of the observer is not high enough to learn about non-perturbative quantum gravity.  Our model is able to show this crossover, but only because it makes sense when $S_{Ob}$ is finite. 

\item \textbf{Meaning of the unique state:}  In this paper we introduced a code $\hat{V}$ that maps the Hilbert space of possible matter states into the ``fundamental'' Hilbert space $\mathcal{H}_{Ob'}$.  If the full Hilbert space is one dimensional however, why do we need this big Hilbert space of possible initial states?  Can't we just say that the universe is always in its unique state?  For example in the code model from figure \ref{Ocodefig} the unique state is
\be
|HH\ran_{eff}=O^T|0\ran.  
\ee
We call this the Hartle-Hawking state because if we compute the inner product of other states in the (observerless) code, dropping $|\psi_0\ran$ to make sure the Hartle-Hawking state is an allowed input, we get 
\be
\lan \phi|V^\dagger V|\psi\ran=\frac{1}{d}\lan \phi|V^\dagger V|HH\ran\lan HH|V^\dagger V|\psi\ran,
\ee
which is one way of defining the Hartle-Hawking state \cite{Marolf:2020xie,McNamara:2020uza,Usatyuk:2024isz}.  The issue however is that even if we start in the Hartle-Hawking state, as soon as the observer starts looking at the universe around them they start projecting onto other states in the effective description.  This kind of conditioning is essential in quantum cosmology, for example it is the reason that scientists in Europe and Asia agree on the pattern of CMB fluctuations even though these are fundamentally quantum fluctuations from inflation.  The more the observer learns about the universe, the more projections they apply to the Hartle-Hawking state in constructing their picture of the universe.  This is completely consistent with the point of view that really the observer together with the rest of the universe is still just in the Hartle-Hawking state.  Our theory is for use by an observer who has already learned the basic facts about themself and their immediate environment, and that observer has a sizable Hilbert space of states to choose from. The entanglement in the state $|\omega\ran$ represents the ``residual'' entanglement between the observer and the universe which they have not removed by doing this conditioning.   

\item \textbf{Nonlinear observables:} So far in this paper we have used the approximate preservation of the inner product by the code $\hat{V}$ as a stand-in for having a theory of measurement for an observer in a closed universe.  For observables $X$ of low rank in the effective description such as $|i\ran\lan i|_M$ this is clearly sufficient, as we can simply define $\wt{X}=VXV^\dagger$ and then we have
\be
\lan \phi|\hat{V}^\dagger\wt{X}\hat{V}|\psi\ran\approx \lan \phi|X|\psi\ran,
\ee
with the approximation holding up to an error which is exponentially small in $S_{Ob}$ (or possibly $S_0$ if $S_{Ob}\gg S_0$).  Low-rank observables are rather unusual however, a more local observable, e.g. such as a single-site Pauli matrix, has large rank.  For these we must in general use the non-linear state-specific reconstruction of \cite{Akers:2021fut,Akers:2022qdl}.\footnote{The exception is when the matter Hilbert space is small compared to $S_2(\omega_{Ob})$, since then $\hat{V}$ is an approximate isometry and we can use $\wt{X}$ for all $X$.}  This is a bit easier to explain if we restrict to pure initial states of $M$.  By the methods of \cite{Akers:2022qdl}, the encoding map $\hat{V}$ is approximately invertible if we restrict to pure states whose complexity is sub-exponential in $S_{Ob}$.  This inverse is not a linear map however, as the set of sub-exponential states is not a linear subspace.  Given a pair of states $|\wt{\psi}\ran,|\wt{\phi}\ran\in \mathcal{H}_{Ob'}$ which we know are the images of sub-exponential states, we thus define the expectation value of $X$ in these states as
\be
\lan X\ran_{\wt{\phi},\wt{\psi}}=V^{-1}(\lan\wt{\phi|})XV^{-1}(|\wt{\psi}\ran).
\ee
By the approximate preservation of the inner product we then have
\be
\lan X\ran_{\wt{\phi},\wt{\psi}}\approx \lan \phi|X|\psi\ran
\ee
up to errors which are exponentially small in $S_{Ob}$ (or again possibly $S_0$ of $S_{Ob}<S_0$), so this measurement theory agrees with semiclassical expectations.

\item \textbf{Importance of negative $\Lambda$:} In various places in this paper we used gravitational theories with negative cosmological constant, in particular for our calculations in JT gravity and for the microscopic model of section \ref{ETHsec}.  We therefore would like to emphasize that the basic idea of this paper, meaning the code shown in figure \ref{Ocode2fig} as a quantum gravity theory for an observer in a closed universe, does \textit{not} rely on having negative cosmological constant or a CFT holographic dual.  We are hopeful that this idea will also be present in a microscopic theory of de Sitter space, one promising indication in this direction is the recent demonstration that the scrambling time in de Sitter space grows with the logarithm of the observer mass \cite{Kolchmeyer:2024fly}.  
\ei

\paragraph{Acknowledgments:}  We thank Daine Danielson, Luca Iliesiu, David Kolchmeyer, Adam Levine, Juan Maldacena, Gautam Satishchandran, Douglas Stanford, Lenny Susskind, and Bob Wald for helpful discussions. DH is supported by the Packard Foundation as a Packard Fellow, the US Department of Energy under grants DE-SC0012567 and DE-SC0020360, and the MIT department of physics. MU was supported in part by grant NSF PHY-2309135 to the Kavli Institute for Theoretical Physics (KITP), and by a grant from the Simons Foundation (216179, LB). YZ was supported in part by the National
Science Foundation under Grant No. NSF PHY-1748958, by a grant from the Simons Foundation (815727, LB), by Heising-Simons Foundation award-6950153, and by David and Lucile Packard Foundation award-2749255.

\appendix
\section{Orthogonal integrals}\label{Oapp}
\bfig
\includegraphics[height=7cm]{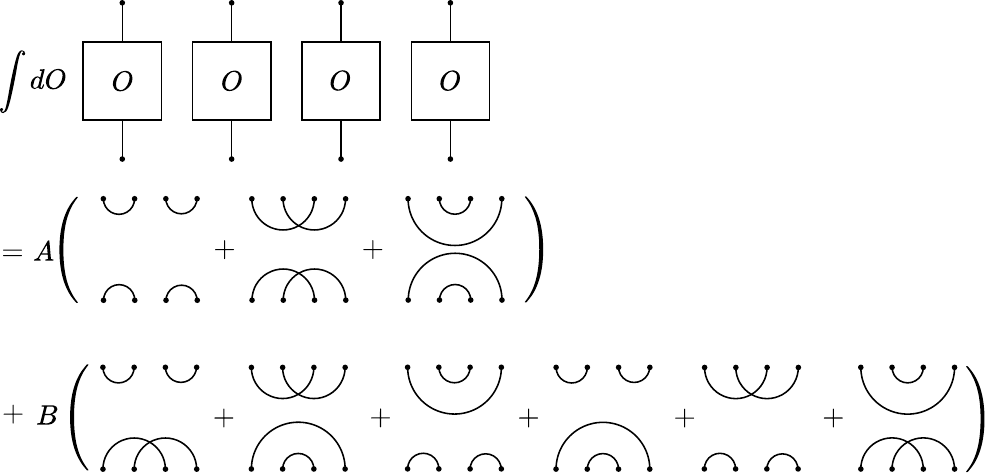}
\caption{Computing the Haar integral of four powers of an orthogonal matrix using pairings and Weingarten coefficients.}\label{Ointfig}
\efig
In this appendix we review how to compute moments of the Haar measure on the orthogonal group $O(d)$.  By rotational invariance these can be decomposed as a sum over pairings of Kronecker $\delta$ symbols:
\be
\int dO\, O_{i_1 j_1}\ldots O_{i_{2k},j_{2k}}=\sum_{P,S}W_{P,S}\prod_{(i,i')\in P}\delta_{ii'}\prod_{(j,j')\in S}\delta_{jj'},
\ee
where $P$ and $S$ are unordered pairings of $2k$ objects (the integral vanishes if $k$ is a half integer) and $W_{P,S}$ are called the orthogonal Weingarten coefficients.  The Weingarten coefficients are invariant if we simultaneously permute the two pairings (as this just permutes the $O$ matrices), so they depend only on the relative permutation between the two pairings $P$ and $S$.  There are general representation-theoretic formulas available for them \cite{Collins:2006jgn}, but for us it will be enough to understand the cases $k=1,2$.  For $k=1$ we simply have
\be
\int dO \, O_{i_1j_1}O_{i_2 j_2}=\frac{1}{d}\delta_{i_1 i_2}\delta_{j_1j_2},\label{1O}
\ee
as can be easily confirmed by noting that if we contract $j_1$ and $j_2$ then we must get $\delta_{i_1 i_2}$ (we are normalizing the Haar measure so that $\int dO=1$).  For $k=2$ there are nine terms; it is easier to draw a picture than to write the equation so see figure \ref{Ointfig}.  The Weingarten coefficients which appear are
\begin{align}\nonumber
A&=\frac{d+1}{d(d+2)(d-1)}\\
B&=-\frac{1}{d(d+2)(d-1)}, 
\end{align}
as can be confirmed without much difficulty by contracting a pair of $j$ indices in figure \ref{Ointfig} to remove two of the $O$ matrices and  using \eqref{1O}.  Note that the ``crossed'' pairings where $P\neq S$ are suppressed by an extra factor of $d$, so typically when we write large $d$ formulas we only need to include the ``diagonal'' pairings where $P=S$ (the possible exception to this rule, which does not actually arise in any of our calculations, is when an index contraction produces a factor of $d$ in the numerator which cancels this suppression).

\bfig
\includegraphics[height=5.2cm]{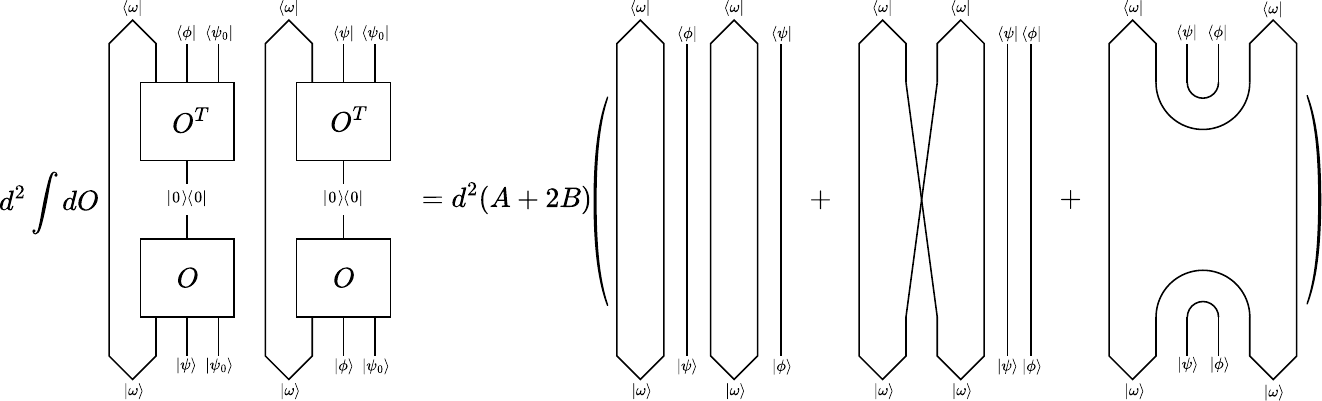}
\caption{Computing a code integral with four $O$s.}\label{VOfig}
\efig
As an illustration of this method, we'll give the details for how to compute the quantity
\be
\int dO |\lan\phi|\hat{V}^\dagger\hat{V}|\psi\ran|^2
\ee
in the code from figure \ref{Ocode2fig}, which feeds into the first line of equation \eqref{codeipobsresult}.  We represent this quantity graphically in the left side of figure \ref{VOfig}.  We can then evaluate the integral by pairing the ingoing and outgoing indices of the $O$s using figure \ref{Ointfig}.  There is a very convenient simplification in that the outgoing indices are all just projected onto $|0\ran$, so the outgoing pairing doesn't matter and we can combine the terms in figure \ref{Ointfig} in groups of three.   Thus we only need to sum over the three pairings of the ingoing indices and multiply by
\be
d^2(A+2B)=\frac{d(d+1)}{(d+2)(d-1)}\left(1-\frac{2}{d+1}\right)=\frac{d}{d+2},
\ee
as shown in the right side of figure \ref{Ointfig}.  Evaluating the right side we have
\be
\int dO |\lan\phi|\hat{V}^\dagger\hat{V}|\psi\ran|^2=\frac{d}{d+2}\Bigg[|\lan\phi|\psi\ran|^2+e^{-S_2(\omega_{Ob'})}+\Tr\left(\omega_{Ob}\omega_{Ob}^T\right)|\lan\phi^*|\psi\ran|^2\Bigg],
\ee
as quoted in equation \eqref{codeipobsresult}.  The other orthogonal integrals quoted in the main text can be evaluated in exactly the same way.

\section{A more structured code model} \label{app:structuredcode}
In the code defined by figure \ref{Ocode2fig} we can choose whatever orthogonal matrix $O$ that we like. In section \ref{sec:code_generic} we studied case of generic $O$ in the Haar measure, and saw that the code preserves a non-trivial Hilbert space whose dimension is upper bounded by the observer entropy $e^{S_{Ob'}}$. This choice can reproduce the gravity results  \eqref{eq:Renyi_topological}, \eqref{eq:JT_entropy} and the ETH result \eqref{eq:entropy_ETH} when the observer entropy is small compared to $S_0$.  In this appendix we will study a different choice of $O$, which reproduces all leading terms in the gravity and ETH calculations when parameters are appropriately identified.

\begin{figure}
  \centering
  \includegraphics[width = 3.6in]{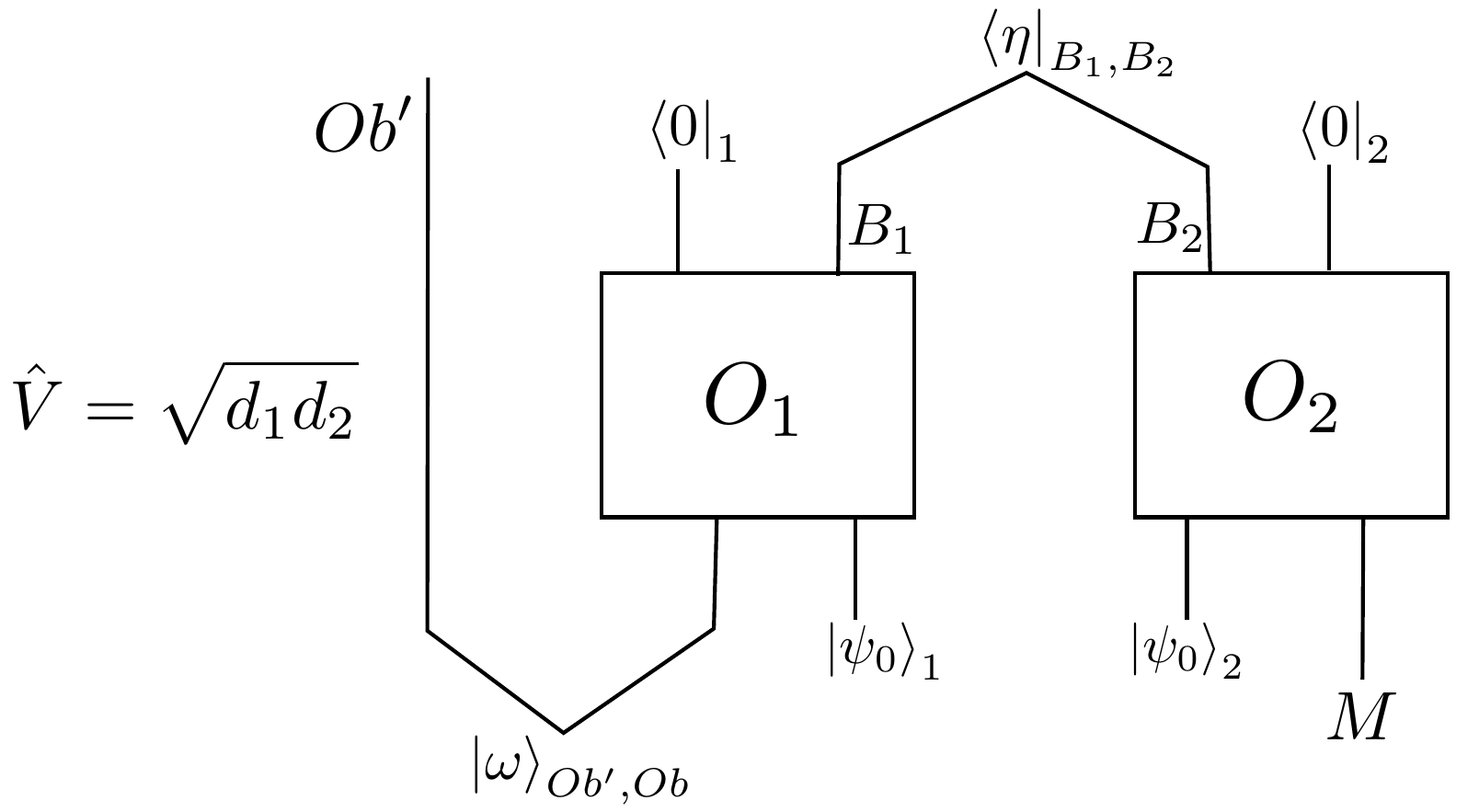}
  \caption{$d_1$ is the dimension of orthogonal matrix $O_1$ and $d_2$ is the dimension of $O_2$. The prefactor $\sqrt{d_1d_2}$ is chosen such that on average the inner product is preserved$\int dO_1 d O_2\bra{\phi}\hat V^{\dagger}\hat V\ket{\psi} = \bra{\phi}\ket{\psi}$.
}
  \label{fig:code_1}
\end{figure}

We give $O$ more structure by separating it into two parts $O_1$ and $O_2$ (Figure \ref{fig:code_1}), with a post selection $\bra{\eta}$ connecting them. We let $O_1$ and $O_2$ be generic orthogonal matrices, and take the state $\ket{\eta}$ to be $\CRT$-invariant. 
With this choice of code, the fluctuation of the inner product is given by
\begin{align}
\label{eq:code_structured_1}
&\int dO_1 dO_2\qty|\bra{\phi}\hat V^{\dagger}\hat V\ket{\psi}-\bra{\phi}\ket{\psi}|^2\nonumber\\
=\ &\qty|\bra{\phi}\ket{\psi}|^2\qty[e^{-S_2(\omega_{Ob})}e^{-S_2(\eta_{B_1})}+\Tr(\omega_{Ob}\omega_{Ob}^T)e^{-S_2(\eta_{B_1})}]\nonumber\\
&\ +\qty|\bra{\phi^*}\ket{\psi}|^2\qty[e^{-S_2(\eta_{B_1})}+\Tr(\omega_{Ob}\omega_{Ob}^T)+e^{-S_2(\eta_{B_1})}e^{-S_2(\omega_{Ob})}]\nonumber\\
&\ +e^{-S_2(\eta_{B_1})}+e^{-S_2(\omega_{Ob})}+\Tr(\omega_{Ob}\omega_{Ob}^T)e^{-S_2(\eta_{B_1})}+\mathcal{O}\qty(\frac{1}{d})
\end{align}
At infinite $d$, there are nine ways to contract the $O$'s in computing the average of $\lan \phi|\hat{V}^\dagger\hat{V}|\psi\ran$, with one of them being removed when we subtract $|\lan\phi|\psi\ran|^2$ to compute the fluctuation.  These precisely correspond to the nine different ways of contracting the $R$'s in ETH calculation of \eqref{eq:ETH_average_1}. 
Imposing $\CRT$ invariance on $\omega$ and $\eta$, \eqref{eq:code_structured_1} simplifies to
\begin{align}
\label{eq:code_structured}
   & \int dO_1 dO_2\qty|\bra{\phi}\hat V^{\dagger}\hat V\ket{\psi}-\bra{\phi}\ket{\psi}|^2\nonumber\\
   =\ &\qty(e^{-S_2(\omega_{Ob})}+e^{-S_2(\eta_{B_1})})\qty(1+\qty|\bra{\phi}\ket{\psi}|^2)+e^{-S_2(\eta_{B_1})}e^{-S_2(\omega_{Ob})}\qty(1+3\qty|\bra{\phi}\ket{\psi}|^2)+\mO\qty(\frac{1}{d})
\end{align}
We can again entangle the matter with a reference system $\chi$ and consider its second Renyi entropy, we obtain
\begin{align}
\label{eq:code_structured_Renyi}
   & \int d O_1 dO_2\Tr(\Psi_{M'}^2)\nonumber\\
=\ &\Tr(\omega_{Ob'}^2)+\Tr(\chi_{M'}^2)+\Tr(\chi_M\chi_M^T)\Tr(\omega_{Ob}\omega_{Ob}^T)\nonumber\\
&\ \ +e^{-S_2(\eta_{B_1})}\Big[1+\Tr(\omega_{Ob}\omega_{Ob}^T)+\Tr(\chi_M\chi_M^T)\nonumber\\
&\ \ \ \ \ +\Tr(\chi_M^2)\Tr(\omega_{Ob}^2)+\Tr(\chi_M^2)\Tr(\omega_{Ob}\omega_{Ob}^T)+\Tr(\omega_{Ob}^2)\Tr(\chi_M\chi_M^T)\Big]
\end{align}
Note that \eqref{eq:code_structured},\eqref{eq:code_structured_Renyi} match exactly with the ETH calculations in \eqref{eq:fluctuation_ETH},\eqref{eq:Renyi_ETH} once we identify $\frac{Z_2}{Z_1^2}$ with $e^{-S_2(\eta_{B_1})}$.  The potentially leading terms in \eqref{eq:code_structured},\eqref{eq:code_structured_Renyi}, meaning the three terms involving only one of the entropies, also agree with the gravitational expressions \eqref{topips0} and \eqref{eq:Renyi_topological_2}.

\section{Replica wormholes and extremal surfaces in closed universe}
\label{app:RT}
In this appendix we give additional technical details on the JT gravity calculations in the main text. 

We want to sum over geometries with $2n$ asymptotic boundaries with some number of matter operator insertions on each boundary. The Euclidean JT gravity path integral minimally coupled to QFT matter localizes onto hyperbolic geometries, with the sum over geometries accompanied by a sum over all geodesics connecting operator insertions of the same flavor.\footnote{With QFT matter (as opposed to worldline matter), there is an additional sum over all closed geodesics on the geometry. We will ignore this subtlety.} Each geodesic gives a contribution $\exp(-\Delta \ell)$ where $\Delta$ is the AdS scaling dimension related to the mass by $m^2=\Delta(\Delta-1)$ and $\ell$ is the length of the geodesic, appropriately renormalized if the geodesic is anchored to an asymptotic boundary. Unfortunately, the full path integral is divergent once we include geometries where there are geodesic cycles that pinch to zero size $b \to 0$. It is standard to treat JT coupled to matter as an effective theory and only sum over saddlepoints, or a subset of geometries. When the boundary operator insertions have large scaling dimensions there is typically a saddlepoint geometry where none of the cycles are pinched, and therefore the sum over geodesics is convergent. We now explain how to construct the geometries we are interested in section \ref{sec:4.3JT}.

The basic building block is a Euclidean rectangle, with opposing sides having asymptotic boundary conditions with lengths $\beta_L, \beta_R$, connected by geodesics along the top and bottom of lengths $\ell_1, \ell_2$. The full path integral evaluated with these boundary conditions is \cite{Yang:2018gdb,Saad:2019pqd,Stanford:2020wkf,Hsin:2020mfa}
\be
\text{JT rectangle:}\quad =\quad  \int_{0}^\infty d E \rho(E) e^{-(\beta_L +\beta_R) E} \lan \ell_1 | E\ran \lan E | \ell_2 \ran\,,
\ee
with 
\be
\rho(E) = \frac{e^{S_0}}{4\pi^2} \sinh(2\pi \sqrt{E})\,, \qquad \lan \ell | E\ran \equiv h_s(\ell)= 2^{3/2} K_{2is}(2e^{-\ell/2})\,, \qquad E = s^2\,.
\ee
The overlap $\lan E | \ell \ran$ is a disk topology with a geodesic boundary $\ell$ bounding a fixed energy boundary condition \cite{Yang:2018gdb}. The cylinder with one operator inserted on each boundary can be computed by gluing the geodesic boundaries of the rectangle together by setting $\ell_i=\ell$ and inserting a weight for the matter propagator $e^{-S_0}\int_{-\infty}^\infty d \ell e^{-\Delta \ell}$
\begin{align}
\text{Cylinder with matter insertion:} \quad & = e^{-S_0}\quad \int_{0}^\infty d E \rho(E)  e^{-(\beta_L +\beta_R) E}  \int_{-\infty}^\infty  d\ell \lan \ell | E\ran \lan E | \ell \ran e^{-\Delta \ell}\,, \nonumber\\
& = e^{-S_0}\int_0^\infty d E \rho(E) e^{-(\beta_L + \beta_R) E} \frac{2 \Gamma(\Delta \pm i s \pm i s')}{\Gamma(2\Delta)}
\end{align}
The above integral includes all geodesics, including infinitely many winding geodesics that connect two operators on opposite asymptotic boundaries. The integral has been simplified using
\be \label{eqn:Besselintegral}
G_\Delta(E,E')\equiv 2^3\int_{-\infty}^\infty d \ell e^{-\Delta \ell} K_{2 i s}(2 e^{-\ell/2}) K_{2 i s'}(2 e^{-\ell/2}) = \frac{2 \Gamma(\Delta \pm i s \pm i s')}{\Gamma(2\Delta)} \,.
\ee
\begin{figure}
    \centering
    \includegraphics[width=0.4\linewidth]{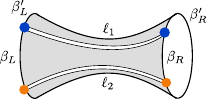}
    \caption{Construction of the geometry $Z_1$ used in the main text. The cylinder with four total operator insertions can be evaluated by gluing rectangular geometries together. In the figure we have cut the cylinder open into two rectangles that are glued along their geodesic boundaries. The geodesic distance between the blue operator insertions is given by $\ell_1$, and the distance between the orange is given by $\ell_2$. }
    \label{fig:Z1app}
\end{figure}
We can now calculate the cylinder amplitude with two operator insertions on each asymptotic boundary given by equation \eqref{eqn:JT_inner_prod}. In the limit that the scaling dimensions of both operator insertions is large, the dominant contribution will come from the shortest geodesic on the cylinder geometry. This is the geodesic that crosses straight through the wormhole. We have two such geodesics as shown in the graphic in \eqref{eqn:JT_inner_prod}. This geometry is obtained by gluing two rectangles together with insertions of $e^{-2S_0}\int d\ell e^{-\Delta_{Ob} \ell} \int d\ell e^{-\Delta_{M} \ell'}$, see figure \ref{fig:Z1app}. Using \eqref{eqn:Besselintegral} we get\footnote{This answer implicitly includes a sum over winding geodesics that connect the operators such that the two distinct geodesics never intersect. There are also contributions where the geodesics connecting the blue and orange operator insertions intersect, but these are not included in the above formula and they are suppressed when the scaling dimension is large.}
\be \label{eqn:Z1}
Z_1 \equiv e^{-2 S_0} \int_0^\infty d E d E' \rho(E) \rho(E') e^{-(\beta_L +\beta_R )E - (\beta_L' + \beta_R') E'} G_{\Delta_{Ob}}(E,E') G_{\Delta_{M}}(E,E')\,.
\ee
This integral has a saddlepoint value in the energies $E_1, E_2$.

For Renyi entropies and the von Neumann entropy we will need geometries with $2n$ asymptotic boundaries with two operator insertions on each boundary. The two important sets of geometries are those where all boundaries connect in mutual pairs as cylinders, and where all boundaries are fully connected. The fully connected geometries can be built by gluing together two pinwheel geometries. A pinwheel geometry has $2n$ segments of asymptotic boundaries of length $\beta_i$ interspersed with $2n$ geodesic boundaries of lengths $\ell_i$ between each asymptotic boundary. The exact expression for the pinwheel is \cite{Penington:2019kki,Hsin:2020mfa}
\be \label{eqn:planar_geometry}
\int_0^\infty d E \rho(E) \prod_{i=1}^{2n} e^{-\beta_i E} \lan E | \ell_i \ran\,.
\ee
The amplitude for the fully connected geometry is given by gluing two pinwheels together along their geodesic boundaries by inserting $\prod_{i=1}^{2n} e^{-S_0}\int_{-\infty}^\infty d\ell_i e^{-\Delta_i \ell_i}$ between two copies of \eqref{eqn:planar_geometry} where $\Delta_i$ determines the scaling dimension of the operator insertions on the respective asymptotic boundary. 

Specializing to the case of interest in the main text, we take the geodesics to be alternating and weighed by $\Delta_{Ob}$ and $\Delta_M$. We also take each asymptotic boundary to have operator insertions separated by $\beta,\beta'$. Using the integral identity \eqref{eqn:Besselintegral} we find the answer for the pinwheel to be
\be \label{eqn:Znapp}
 Z_n \equiv e^{-2n S_0}\int_0^\infty d E d E'\rho(E)\rho(E') \left( e^{-\beta E-\beta' E'}G_{\Delta_{Ob}}(E, E') G_{\Delta_M}(E, E')\right)^n\,.
\ee
We have that the connected pinwheel geometry scales with the topological factor as $Z_n \sim e^{2S_0(1-n)}$. Figure \ref{fig:2ndrenyi_JT} show the $Z_{n=2}$ geometry obtained by gluing two pinwheels. 

We can also compute the vN entropy in the interesting regime where the fully connected pinwheel geometry dominates. This corresponds to the two quantum extremal surfaces in figure \ref{fig:jt_closed_universe} dominating the entropy. The entropy of the matter reference system in this case is given by
\begin{align}
	S_{vN} = \lim_{n\rightarrow 1}\frac{1}{1-n}\log(\frac{\Tr\rho^n}{(\Tr\rho)^n}) = \lim_{n\rightarrow 1}\frac{1}{1-n}\log(\frac{Z_n}{Z_1^n})\,.
\end{align}
We can analytically continue the expression for $Z_n$ to $n\approx 1$. In the saddlepoint approximation the integrals are peaked around energies values $E_{i}^*$. All terms except the density of states cancel in the ratio inside the logarithm giving
\begin{align}
\label{eq:QES_replica}
S_{vN} &=\log\qty[\rho(E_1^*)]+\log\qty[\rho(E_2^*)] \nonumber\\
&= \left.\left( S_0 + 2\pi\Phi \right)\right\rvert_{\text{QES}_1}+ \left.\left( S_0 + 2\pi\Phi \right)\right\rvert_{\text{QES}_2}\,.
\end{align}
where $E_i^*$ is the saddle energy for $n = 1$, obtained by extremizing $Z_1$ in \eqref{eqn:Z1}. The two terms should be interpreted as the value of the dilaton at the two extremal surfaces in the closed universe in figure \ref{fig:jt_closed_universe}. The saddlepoint equations for the energy integrals are exactly the same as the equations giving the classical geometry of the closed universe found in appendix A of \cite{Usatyuk:2024mzs}, as well as the density of states at the saddlepoint energies matching the value of the QES determined there.

The simplest case is where the matter and observer are antipodal to each other given by $\beta=\beta'$ giving a $Z_2$ symmetric closed universe with geodesic size $b$ determined by $\beta, \Delta_{Ob}, \Delta_M$. In this case $E_1^*=E_2^*=E^*$ and the saddlepoint equations simplify 
\be
\sqrt{E^*} \beta = \pi-\text{arctan}\frac{2 \sqrt{E^*} }{\Delta_{M}} - \text{arctan}\frac{2 \sqrt{E^*} }{\Delta_{Ob}}\,,
\ee
where we used the expansion of the gamma functions for large arguments in \eqref{eqn:Z1}. In the case of light matter this saddle can be expanded and the found to match the QES of the dilaton in appendix A of \cite{Usatyuk:2024mzs}.

\section{Matter in presence of a QES and an observer}
\label{app:JT_QES}

In this appendix we compute the the contribution from figure \ref{fig:JT_QES_inner_product_square}(c). 

\subsection{Without matter backreaction}

\begin{figure}
  \centering
  \includegraphics[width = 2in]{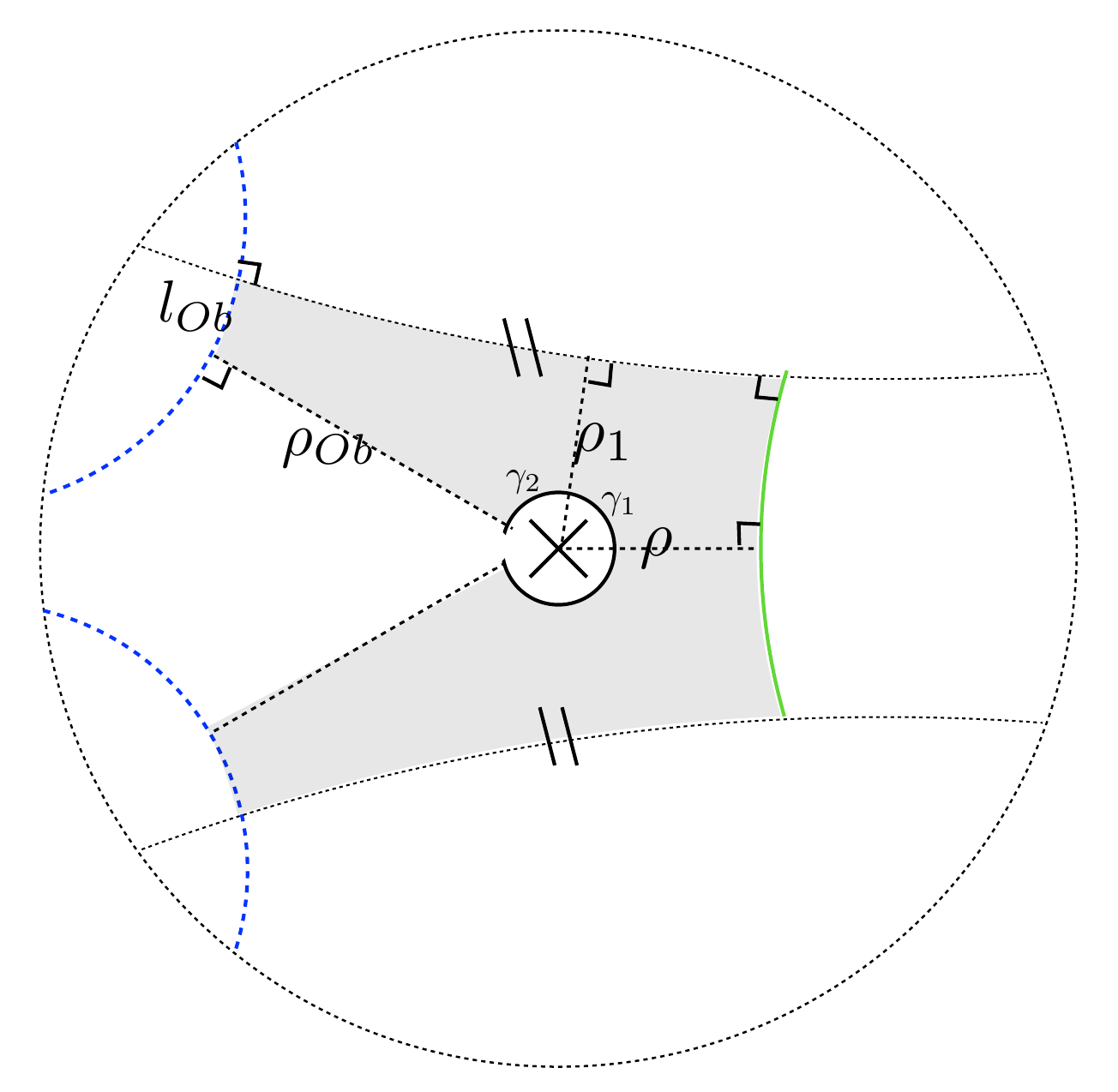}
  \caption{}
  \label{fig:QES_large_0}
\end{figure}

As a warm up, we first evaluate a simpler configuration where we ignore the backreaction from the matter such that the two quantum extremal surfaces coincide. Figure \ref{fig:QES_large_0} is one half of the cylinder cut open.

We will work in disk coordinate where the metric and dilaton are given by
\begin{align}
\label{eq:disk}
	ds^2 = d\rho^2+\sinh^2\rho d\theta^2,\ \ \ \ \Phi  = \Phi_c\cosh\rho
\end{align}
where $\Phi_c$ is the value of the dilaton at the disk center.
The boundary condition near the QES sets $\Phi_c = \Phi_e\sqrt{1-\alpha^2}$, while the boundary condition at the end of the world brane sets 
\begin{align}
	\sinh\rho_{Ob} = \frac{\mu}{\Phi_c},\ \ \ \Phi_{Ob} = \sqrt{\mu^2+\Phi_c^2}
\end{align}

We first evaluate the action with fixed $\rho$ and $\rho_1$ (Figure \ref{fig:QES_large_0}), and then extremize with respect to $\rho_1$ and $\rho$. The length of the end of world brane $2l_{Ob}$ satisfies
\begin{align}
	\sinh l_{Ob} = \frac{\sinh\rho_1}{\cosh\rho_{Ob}}
\end{align}

The angles satisfy
\begin{align}
	\cos\gamma_1 = \tanh\rho_1\tanh\rho,\ \ \ \cos\gamma_2 = \tanh\rho_1\tanh\rho_{Ob}
\end{align}
The action of the entire cylinder with fixed $\rho_1$ and $\rho$ is given by
\begin{align}
	-I_E =\ & 2\qty[\Phi_e\qty(\alpha-\frac{1}{\tanh\rho_e})\sinh\rho_e(2\gamma_1+2\gamma_2)-2\mu l_{Ob}+\pi(\Phi_{Ob}+\Phi_e)]\nonumber\\
	=\ &-4\Phi_c\qty(\gamma_1+\gamma_2+l_{Ob}\sinh\rho_{Ob} )+2\pi(\Phi_{Ob}+\Phi_e)\equiv -4\Phi_c f(\rho, \rho_1)+2\pi(\Phi_{Ob}+\Phi_e)
\end{align}
We see that we need to extremize the following function with respect to $\rho_1$ and $\rho$: 
\begin{align}
f(\rho,\rho_1)\equiv\arccos(\tanh\rho_1\tanh\rho)+\arccos(\tanh\rho_1\tanh\rho_{Ob})+\sinh\rho_{Ob}\arcsinh\qty(\frac{\sinh\rho_1}{\cosh\rho_{Ob}})
\end{align}
The saddle point is at $\rho_1=0$ and $\rho=0$. With finite positive $\alpha$ we need $\rho\geq \rho_e$ and $\rho_1\geq \rho_e$. So the extremal value happens when $\rho=\rho_1 = \rho_e$. When $\alpha\rightarrow 0$, $\rho_e\rightarrow 0$, the extremal action is given by
\begin{align}
	-I_E = 2\pi(\Phi_{Ob}-\Phi_e)
\end{align}

\subsection{With matter backreaction}
\begin{figure}
  \centering
  \includegraphics[width = 2.2in]{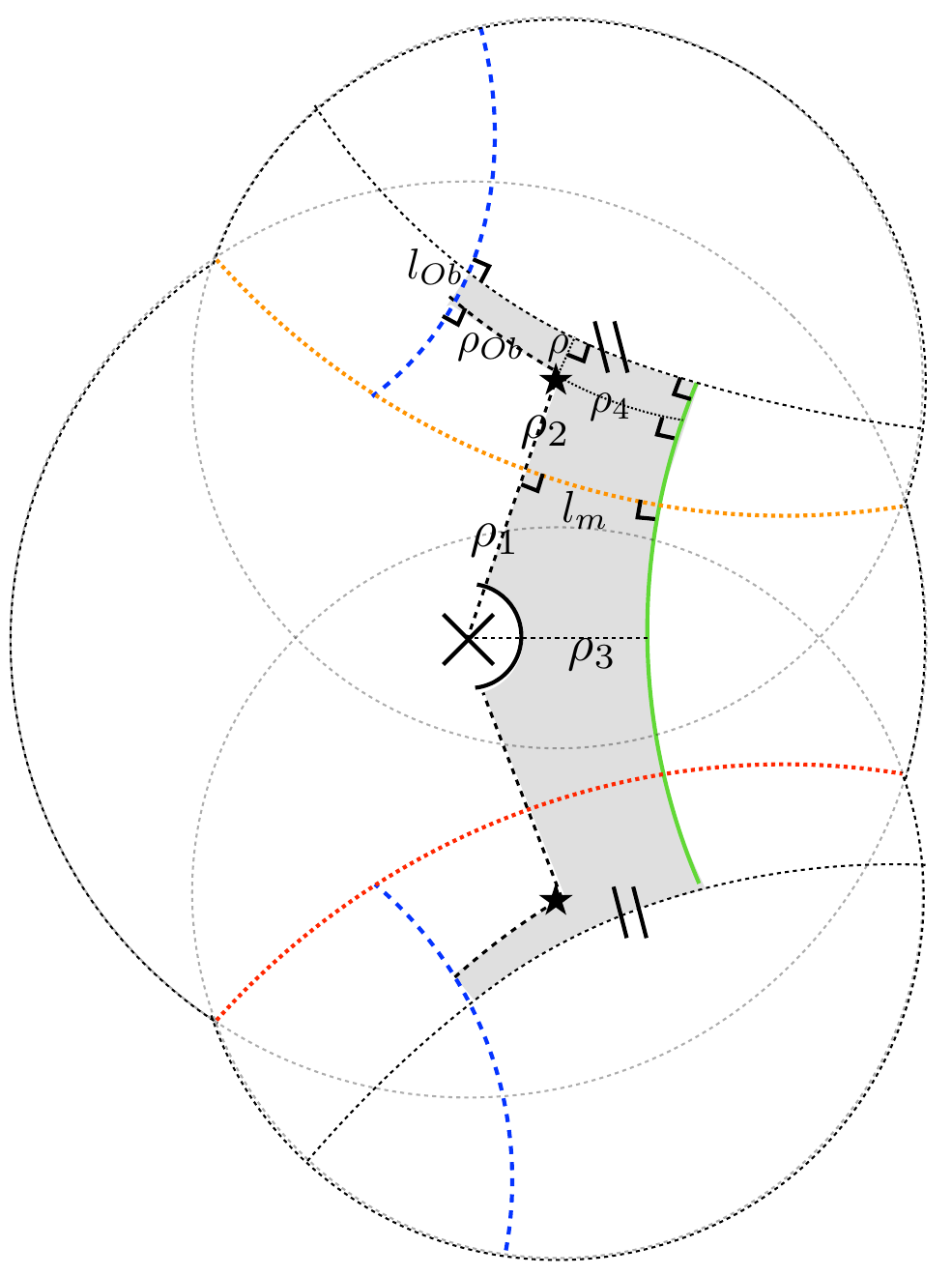}
  \caption{}
  \label{fig:QES_large_1}
\end{figure}

Figure \ref{fig:QES_large_1} is again one half of the cylinder cut open. First, note that figure \ref{fig:QES_large_1} reduces to figure \ref{fig:QES_large_0} when we ignore back reaction and the quantum surfaces coincide. 

We again work in disk coordinate as in \eqref{eq:disk}.
We evaluate the action as a function of fixed $l_m$ and $\rho$ (Figure \ref{fig:QES_large_1}), then extremize with respect to these two variables. 

From the continuity condition at the matter, we have
\begin{align}
	&\Phi_c\cosh\rho_1 = \Phi_{e2}\cosh\rho_2\\
	&\Phi_c\sinh\rho_1+\Phi_{e2}\sinh\rho_2 = m
\end{align}
which implies
\begin{align}
	&\sinh\rho_1 = \frac{1}{2\Phi_c}\qty(m+\frac{\Phi_{e2}^2-\Phi_c^2}{m})\\
	&\sinh\rho_2 = \frac{1}{2\Phi_{e2}}\qty(m-\frac{\Phi_{e2}^2-\Phi_c^2}{m})
\end{align}

From the boundary condition at the end of the world brane, we have
\begin{align}
	\sinh\rho_{Ob} = \frac{\mu}{\Phi_{e2}},\ \ \Phi_{Ob} = \sqrt{\mu^2+\Phi_{e2}^2}
\end{align}

From geometry, we have
\begin{align}
	\sinh \rho_3 = \cosh\rho_1\sinh l_m,\ \ 
	\sinh\rho_4 = \cosh\rho_2\sinh l_m,\ \ 
	\sinh l_{Ob} = \frac{\sinh\rho}{\cosh\rho_{Ob}}
\end{align}

The action of the cylinder is given by
\begin{align}
	-I_E = \ &4\Bigg[\Phi_e\qty(\alpha-\frac{1}{\tanh\rho_e})\sinh\rho_e\arccos(\tanh\rho_1\tanh\rho_3)-ml_m-\mu l_{Ob}\nonumber\\
	&\ \ -\Phi_{e2}\Big(\arccos(\tanh\rho_2\tanh\rho_4)+\arccos(\tanh\rho_4\tanh\rho)+\arccos(\tanh\rho\tanh\rho_{Ob})\Big)\nonumber\\
    &\ \ +\frac{\pi}{2}(\Phi_e+\Phi_{Ob})+\pi\Phi_{e2}\Bigg]\nonumber\\
	\equiv\ &4\qty[-f(\rho,l_m)+\frac{\pi}{2}(\Phi_e+\Phi_{Ob})+\pi\Phi_{e2}]
\end{align}

We need to extremize the following function with respect to $\rho$ and $l_m$:
\begin{align}
	f(\rho,l_m)\equiv\ &\Phi_c\arccos(\frac{\sinh\rho_1\sinh l_m}{\sqrt{1+\cosh^2\rho_1\sinh^2 l_m}})+(\Phi_c\sinh\rho_1+\Phi_{e2}\sinh\rho_2) l_m\nonumber\\
    &+\Phi_{e2}\qty[\arccos(\frac{\sinh\rho_2\sinh l_m}{\sqrt{1+\cosh^2\rho_2\sinh^2 l_m}})+\arccos(\frac{\cosh\rho_2\sinh l_m\tanh\rho}{\sqrt{1+\cosh^2\rho_2\sinh^2 l_m}})]\nonumber\\
&+\Phi_{e2}\qty[\arccos(\tanh\rho\tanh\rho_{Ob})+\sinh\rho_{Ob}\arcsinh\qty(\frac{\sinh\rho}{\cosh\rho_{Ob}})]
\end{align}

The saddle point is at $l_m = 0$ and $\rho = 0$. For finite $\alpha$, we need $\rho_3\geq \rho_e$, and the extremal value is at $\rho = 0$ and $l_m = \arcsinh\qty(\frac{\sinh\rho_e}{\cosh\rho_1})$. In the limit where $\alpha\rightarrow 0$, the action is given by
\begin{align}
	-I_E = 2\pi\qty(\Phi_{Ob}-\Phi_{e2})
\end{align}
as we used in obtaining \eqref{eq:JT_QES_fluctuation}-\eqref{eq:JT_QES_Renyi}.

\bibliographystyle{jhep}
\bibliography{bibliography}
\end{document}